%% file: TaskbenchWAMTA26.tex
\documentclass[runningheads]{llncs}

\usepackage[T1]{fontenc}
\usepackage{microtype}

\usepackage[american]{babel}
\usepackage{listings}
\usepackage{graphicx}
\graphicspath{{figs/}}
\usepackage{color}
\usepackage[colorlinks=true,linkcolor=black,citecolor=black]{hyperref}
\usepackage{epstopdf}
\usepackage{caption}
\usepackage{subcaption}
\usepackage[table, dvipsnames, xcdraw]{xcolor}
\usepackage{longtable}
\usepackage{booktabs}
\usepackage[group-separator={,}, group-minimum-digits=4]{siunitx}
\usepackage{xspace}
\usepackage{array}
\usepackage[normalem]{ulem}
\useunder{\uline}{\ul}{}
\usepackage{orcidlink}
\usepackage{todonotes}
\usepackage{multirow}
\usepackage{float}
\usepackage{enumitem}
\usepackage{makecell}
\usepackage{etoolbox}
\patchcmd{\paragraph}{\itshape}{\bfseries\boldmath}{}{}

\usepackage{diagbox}
\usepackage{amsmath}

\usepackage{mathtools, nccmath}
\usepackage{wrapfig}

\begin{document}

\title{
    Exploring Performance-Productivity Trade-offs\\
    in AMT Runtimes: A Task Bench Study\\
    of Itoyori, ItoyoriFBC, HPX, and MPI
}

\titlerunning{A Task Bench Study of Itoyori, ItoyoriFBC, HPX, and MPI}
\authorrunning{Torben R. Lahnor, Mia Reitz, Jonas Posner, and Patrick Diehl}

\author{
    Torben R. Lahnor\,\inst{1}
    \and
    Mia Reitz\,\inst{1}\,\orcidlink{0009-0000-6188-3693}
    \and
    Jonas Posner\,\inst{2}\,\orcidlink{0000-0002-6491-1626}
    \and
    Patrick Diehl\,\inst{3}\,\orcidlink{0000-0003-3922-8419}
}

\institute{
    University of Kassel, Kassel, Germany\\
    \email{torben.lahnor@student.uni-kassel.de},
    \email{mia.reitz@uni-kassel.de}
    \and Fulda University of Applied Sciences, Fulda, Germany\
    \email{jonas.posner@cs.hs-fulda.de}
    \and Los Alamos National Lab (LANL), Los Alamos, U.S.\
    \email{diehlpk@lanl.gov}
}

\hypersetup{
    pdftitle={{Exploring Performance-Productivity Trade-offs in AMT Runtimes: A Task Bench Study of Itoyori, ItoyoriFBC, HPX, and MPI}},
    pdfsubject={WAMTA26},
    pdfauthor={Torben R. Lahnor, Mia Reitz, Jonas Posner, and Patrick Diehl},
    pdfkeywords={{Asynchronous Many-Task (AMT), Itoyori, HPX, Task Bench}}
}

\maketitle
\setcounter{footnote}{0}
\interfootnotelinepenalty=10000

\sloppy
\begin{abstract}
    \input{00abstract}
\end{abstract}

\input{01introduction}
\input{02background}
\input{03impl}
\input{04experiments}
\input{05relatedwork}
\input{06conclusions}

\section*{Acknowledgement}
{\footnotesize
This research was partially funded by the Deutsche Forschungsgemeinschaft (DFG, German Research Foundation) under project numbers~512078735 and~558599020.

This work was supported by the U.S. Department of Energy through the Los Alamos National Laboratory.
Los Alamos National Laboratory is operated by Triad National Security, LLC, for the National Nuclear Security Administration of U.S. Department of Energy (Contract No. 89233218CNA000001). LA-UR-26-20080}

\bibliography{TaskbenchWAMTA26}
\end{document}

%% file: 00abstract.tex
Asynchronous Many-Task~(AMT) runtimes offer a productive alternative to the Message Passing Interface~(MPI).
However, the diverse AMT landscape makes fair comparisons challenging.
Task Bench, proposed by Slaughter~\textsl{et al.}, addresses this through a parameterized framework for evaluating parallel programming systems.

This work integrates two recent cluster AMTs, Itoyori and ItoyoriFBC, into Task Bench for the first time.
Itoyori employs a Partitioned Global Address Space~(PGAS) model with RDMA-based work stealing, while ItoyoriFBC extends it with future-based synchronization.

We compare Task Bench implementations using MPI, HPX, Itoyori, and ItoyoriFBC, evaluating performance and programmer productivity.
Each implementation reflects the programming style it naturally supports: Bulk-Synchronous Parallel~(BSP) for MPI, parallel loops for HPX, and nested fork-join and future-based cooperation for Itoyori and ItoyoriFBC, respectively.
Performance is assessed across compute-bound kernels, weak scaling, and imbalanced and communication-intensive patterns.
We use application efficiency, \emph{i.e.}, the percentage of maximum performance achieved, and Minimum Effective Task Granularity~(METG), \emph{i.e.}, the smallest task duration before runtime overheads dominate.
Productivity is quantified using Lines of Code~(LOC) and Number of Library Constructs~(NLC).

Our results reveal distinct trade-offs.
The MPI implementation achieves highest efficiency for regular, communication-light workloads but requires verbose, low-level code.
The HPX implementation maintains stable efficiency under load imbalance, yet ranks last in productivity, showing that AMTs do not inherently guarantee improved productivity.
The Itoyori implementation achieves highest efficiency in communication-intensive configurations while leading in productivity.
The ItoyoriFBC implementation exhibits slightly lower efficiency than Itoyori, though its future-based synchronization offers potential for irregular workloads.

%% file: 01introduction.tex
\section{Introduction}\label{sec:01-Introduction}

The increasing complexity of High-Performance Computing~(HPC) cluster hardware demands software models that expose high concurrency while managing data movement efficiently.
While the Message Passing Interface~(MPI)~\cite{MPI} remains the de facto standard for distributed-memory parallelism, its reliance on static partitioning and explicit synchronization challenges dynamic, irregular programs.

Asynchronous Many-Task~(AMT) programming is an alternative in which programmers divide computation into many tasks that may be processed in parallel.
An AMT runtime system employs dynamic load balancing of these tasks and overlaps task processing and communication.
However, the AMT landscape remains fragmented, with diverse execution models and APIs complicating appropriate system selection for specific scientific workloads~\cite{ClaudiaTut,TaxoTasksPPAM17,amtstudy,taskAnalysis20}.

To address this challenge, Slaughter \textsl{et al.} introduced a benchmarking system called Task Bench~\cite{TaskBench}, which generates synthetic task graphs representative of various scientific applications.
By controlling parameters such as dependency structure (e.g., stencil, FFT), granularity, and communication volume, Task Bench isolates runtime-specific characteristics, such as scheduling overhead and communication latency, from application-specific noise.
Task Bench has evaluated several systems, including Legion, Realm, StarPU, and HPX~\cite{TaskBenchCharmHPX}.

In this work, we integrate two recent cluster AMTs, Itoyori~\cite{Itoyori23} and \mbox{ItoyoriFBC}~\cite{MiaItoyoriFBCAMTE24}, into Task Bench for the first time.
Itoyori implements a Partitioned Global Address Space~(PGAS) model caching global memory accesses combined with nested fork-join~(NFJ) task parallelism.
For dynamic load balancing, it employs Random Work Stealing~(RDWS) via RDMA.
ItoyoriFBC is an Itoyori variant employing Future-Based Cooperation~(FBC) instead of NFJ, providing greater flexibility.

We conduct performance and productivity evaluations of the Task Bench implementations of Itoyori and ItoyoriFBC, comparing them with those of MPI and HPX.
MPI serves as the baseline for static, low-overhead execution, while HPX represents a mature, C\texttt{++} standards-compliant AMT.
It is important to note that we compare \emph{specific Task Bench implementations} using each system, not the systems themselves in isolation.
Each implementation reflects both the library's capabilities and the programming style it naturally supports:
MPI with Bulk-Synchronous Parallel~(BSP) execution,
HPX with parallel loop abstractions,
and Itoyori/ItoyoriFBC with nested fork-join and future-based cooperation, respectively.
We employ application efficiency and Minimum Effective Task Granularity~(METG) to quantify performance across compute-bound kernels, weak scaling, load imbalance, and communication-intensive patterns.
We further assess programmer productivity through Lines of Code~(LOC) and Number of Library Constructs~(NLC) metrics.
Our evaluation reveals that while MPI dominates in regular workloads, Itoyori achieves superior efficiency in communication-intensive scenarios due to its PGAS cache.
We also observe that HPX maintains robust performance under imbalance but incurs higher overheads for fine-grained tasks.
Moreover, we show that the Itoyori programming model reduces code complexity by nearly 50\% compared to MPI.
Our contributions are:
\begin{itemize}
    \item We propose the first Itoyori and ItoyoriFBC Task Bench implementations and conduct a comparative analysis using them.
    \item We identify and resolve performance bottlenecks in the existing HPX Task Bench implementation~\cite{TaskBenchCharmHPX} to ensure fair comparison.
\end{itemize}

The remainder of this paper is organized as follows:
First, Section~\ref{sec:02-Background} provides background on MPI, HPX, Itoyori, and ItoyoriFBC.
Section~\ref{sec:03-impl} details the Task Bench implementations.
Then, Section~\ref{sec:04-experiments} presents and discusses our experiments.
Finally, Sections~\ref{sec:06-Related-Work} and~\ref{sec:07-Conclusions} cover related work and conclusions, respectively.

%% file: 02background.tex
\section{Background}\label{sec:02-Background}

\textbf{MPI} is the de facto standard for programming clusters.
Traditionally, MPI programs run one process per core which communicate through explicit send and receive operations.
While MPI enables highly optimized code, it lacks support for dynamic load balancing.

\textbf{HPX}~\cite{HPX} is a C\texttt{++} Standard Library-compliant AMT.
Unlike MPI, HPX typically runs multiple threads within a single process per node.
It provides a global address space and supports dynamic load balancing among the threads.

\textbf{Itoyori}~\cite{Itoyori23} is an AMT that integrates PGAS with NFJ.
Itoyori employs a child-first execution policy, branching into newly spawned tasks immediately to maximize cache locality.
Global memory accesses are cached per process and use a checkout/checkin API to ensure memory coherence.

\textbf{ItoyoriFBC}~\cite{MiaItoyoriFBCAMTE24} extends Itoyori by replacing the NFJ programming model with FBC.
In FBC, tasks communicate through so-called \textit{futures}, which represent task results that will eventually be computed and can be passed to other tasks as parameters.
Tasks \textit{touch} a future, which suspends the task until the result has been computed.
This approach enables the expression of dependencies as Directed Acyclic Graphs~(DAGs), making it suitable for irregular applications where the task structure is not strictly nested.

\textbf{Task Bench}~\cite{TaskBench} is designed to evaluate runtime performance across diverse scenarios.
It generates synthetic task graphs where vertices represent tasks and edges represent data dependencies.
The graph structure is configurable via parameters such as \texttt{width} (parallelism), \texttt{steps} (depth), and \texttt{type} (dependency pattern).
We use three graph types: \texttt{stencil} (nearest-neighbor dependencies), \texttt{spread} (long-range dependencies), and \texttt{all\_to\_all} (dense dependencies).

Tasks execute a kernel function configured as either \texttt{compute\_bound} (fixed FLOPs) or \texttt{load\_imbalance} (randomized iterations).
Task Bench allows to evaluate performance using two metrics:
\begin{itemize}
    \item \textbf{Application Efficiency:} The ratio of achieved FLOP/s to the theoretical peak performance of the hardware.
    \item \textbf{METG:} The smallest task duration at which the system maintains 50\% efficiency.
    This metric isolates runtime overheads, as systems with lower overheads can sustain efficiency at finer granularities.
\end{itemize}
Moreover, it allows weak scaling evaluation, where the problem size (graph width) increases proportionally with the number of nodes.

%% file: 03impl.tex
\section{Implementation}\label{sec:03-impl}

For the implementations, we use the official MPI implementation, an existing HPX implementation that we improved, and our new Itoyori and ItoyoriFBC implementations.
All implementations are available online~\footnote{\url{https://doi.org/10.5281/zenodo.18556065}}.

Table~\ref{tab:program-metrics} summarizes productivity metrics for each implementation.
Lines of Code~(LOC) counts non-comment, non-blank lines, while Number of Library Constructs~(NLC) counts distinct API calls such as communication primitives and synchronization constructs.
The high LOC and NLC for HPX reflects that our HPX implementation uses explicit MPI calls for inter-node communication.
Note that the recommended HPX approach to avoid calling MPI explicitly would result in lower numbers.
In contrast, Itoyori's \texttt{checkout}/\texttt{checkin} API encapsulates data movement, reducing library interactions.

\input{04table1}

\subsection{MPI}\label{subsec:mpi-impl}
The MPI implementation follows the standard Task Bench reference architecture~\cite{TaskBench} using a Bulk-Synchronous Parallel~(BSP) programming style.
It employs a static partitioning strategy that decomposes the task graph into vertical strips.
Each process computes different vertical strips, and execution proceeds in bulk-synchronous manner.
In each time step, a process exchanges boundary data with logical neighbors via non-blocking pairwise communication (\texttt{MPI\_Isend} and \texttt{MPI\_Irecv}) followed by a barrier (\texttt{MPI\_Waitall}).
This design minimizes runtime overhead, but performance is limited by the slowest process at each barrier.

\subsection{HPX}\label{subsec:hpx-impl}
The HPX implementation extends the above MPI implementation, while using MPI for inter-node communication and HPX for intra-node parallelism and dynamic scheduling.
One could use HPX without explicitly calling MPI, but we built upon the existing implementation from~\cite{TaskBenchCharmHPX}, which uses MPI for inter-node communication.
We introduced the following improvements:
\begin{itemize}
    \item The original tagging scheme was insufficient for the graph sizes evaluated in this study (see Section~\ref{sec:04-experiments}).
    We refactored the tagging logic to use the maximum tag size ($2^{22}$) of the environment, enabling experiments with up to a width of~32 per core.
    \item The default chunk size for the \texttt{hpx::for\_loop} executor was too large, causing a load imbalance.
    We adjusted the chunking strategy to enable fine-grained dynamic load balancing (chunk size of 1).
    \item We resolved an issue that disabled an optimized communication path for stencil patterns.
    The optimized path avoids unnecessary serialization for intra-node dependencies.
\end{itemize}

\subsection{Itoyori}\label{subsec:itoyori-impl}
Our Itoyori implementation employs a nested fork-join~(NFJ) programming style combined with PGAS-based data access.
We use the PGAS cache to minimize data movement overhead.
We employ two global data structures to store task outputs, alternating their roles (input/output) between time steps to eliminate redundant copying.
The computation begins with a root task that spawns child tasks for the current time step.
Similar to the MPI and HPX implementations, the Itoyori implementation contains a barrier between time steps.

\subsection{ItoyoriFBC}\label{subsec:itoyorifbc-impl}
Our ItoyoriFBC implementation employs a Future-Based Cooperation~(FBC) programming style.
Instead of global memory accesses, the ItoyoriFBC implementation represents task outputs as futures and passes them to the dependent tasks.
Naturally, this would allow tasks of the next time step to execute before finishing all tasks of the current time step.
To enable a fair comparison, we intentionally employ a barrier between time steps similar to the other implementations.
Without barriers, ItoyoriFBC could overlap computation across time steps, potentially improving efficiency by hiding communication latency behind useful work.
However, this would confound the comparison, as the other implementations do not support cross-step overlap.

%% file: 04table1.tex
\begin{table}[ht!]
    \centering
    \begin{tabular}{lcccc}
        \toprule
        & \textbf{MPI} & \textbf{HPX} & \textbf{Itoyori} & \textbf{ItoyoriFBC} \\
        \midrule
        LOC & 137 & 224 & 77  & 115 \\
        \hline
        NLC &  11 &  23 & 14  &  11 \\
        \bottomrule
    \end{tabular}
    \caption{LOC and NLC values for Task~Bench implementations}
    \label{tab:program-metrics}
    \vspace{-1cm}
\end{table}

%% file: 04experiments.tex
\section{Experiments}\label{sec:04-experiments}

We conducted experiments on the Goethe-NHR~\cite{Goethe} supercomputer.
Each compute node has two Intel Xeon Gold 6148 (Skylake) processors (40 cores total) interconnected via InfiniBand.
We used GCC~11.4.1 with Open MPI~5.0.5, HPX~v1.11.0, Itoyori~v0.0.2, and ItoyoriFBC from~\cite{MiaItoyoriFBCAMTE24}.
HPX was configured with the Task Bench build script~\cite{TaskBench} and the MPI parcelport for inter-node communication, using the same Open MPI version as the pure MPI implementation.
Both Itoyori and ItoyoriFBC also use MPI as inter-node communication.
Note that, while HPX provides multiple parcelports, this paper is limited to the MPI parcelport.
We report results as averages over five runs.
We selected graph patterns that represent distinct communication characteristics: \texttt{stencil} for localized, regular communication typical of structured grids; \texttt{spread} for distributed dependencies across the task graph width; and \texttt{all\_to\_all} for worst-case, fully-connected communication.
These patterns cover the spectrum from communication-light to communication-intensive workloads; FFT and tree-based patterns exhibit similar dependency structures and remain candidates for future evaluation.

\subsection{Varying Width and METG}\label{subsec:varying-width}
We first characterize the impact of task graph width on runtime efficiency and determine the METG.
As shown in Figure~\ref{fig:varying-width}, the Itoyori and ItoyoriFBC implementations exhibit reduced efficiency at low task graph widths (\emph{e.g.}, one task per core).
This reduction stems from the RDWS algorithm, which incurs latency penalties.
Profiling reveals that failed steal attempts cost approximately $3\,\mu s$, becoming the dominant factor for fine-grained workloads.
However, increasing the width to 16 tasks per core significantly amortizes this overhead, improving the efficiency of the Itoyori implementation from 25\% to 76\% at task-size $2^{14}$.
The MPI and HPX implementations, using static partitioning and intra-node sharing, respectively, maintain higher efficiency at lower widths but show less relative improvement as width increases.
The METG(50\%) metric confirms this finding: the Itoyori implementation requires coarser task granularity to mask runtime overheads compared to MPI and HPX, though the gap narrows substantially as available parallelism increases.
The HPX implementation results are limited to a width of 32 per core due to MPI tag-space constraints in its inter-node communication layer.

\input{04fig1}

\subsection{Weak Scaling}\label{subsec:weak-scaling}
We focus on weak scaling because Itoyori's RDWS algorithm incurs overhead at low task-per-core ratios (as shown in Section~\ref{subsec:varying-width}), making strong scaling scenarios inherently unfavorable for work-stealing-based systems.

As shown in Figure~\ref{fig:weak-scaling}, the MPI and HPX implementations demonstrate excellent weak scaling, with METG(50\%) remaining stable for HPX and increasing only marginally for MPI.
The localized communication structure of the \texttt{stencil} pattern favors these implementations, as most dependencies are resolved intra-node.
In contrast, the Itoyori and ItoyoriFBC implementations exhibit a performance decline when scaling to multiple nodes.
This decline results from randomized task distribution; even within a stencil pattern, processes frequently access data from remote nodes, incurring higher communication latencies compared to the static mapping of the MPI implementation.

\input{04fig2}

\subsection{Load Imbalance}\label{subsec:imbalance}
We evaluate runtime resilience to irregular workloads using the \texttt{load\_imbalance} kernel.
As expected, the MPI implementation's performance degrades linearly with imbalance (see Figure~\ref{fig:imba}), dropping to $\approx$85\% efficiency at the highest factor due to its lack of dynamic load balancing.
However, its low baseline overhead keeps it competitive.
The HPX, Itoyori, and ItoyoriFBC implementations successfully mitigate imbalance through work stealing, maintaining superior efficiency at high imbalance factors.

\input{04fig3}

\subsection{Communication-Intensive Benchmarks}\label{subsec:communication-focused-benchmarks}
To assess performance under high-bandwidth requirements, we use the \texttt{spread} and \texttt{all\_to\_all} graph patterns.

The \texttt{spread} graph enforces dependencies across the entire width of the task graph.
As shown in Figure~\ref{fig:spread}, the Itoyori implementation achieves high efficiency (>89\% for 40 dependencies).
The MPI implementation performs well initially but degrades as node count increases.
The HPX implementation exhibits a marked performance reduction when scaling (from 87\% to 65\% with 2 nodes).
We suspect that this is due to the static partitioning among nodes.
This result highlights that AMT performance is sensitive to communication topology and underscores the need for decentralized communication.

The \texttt{all\_to\_all} pattern represents the worst-case communication scenario, where every task depends on all tasks from the preceding step.
Figure~\ref{fig:ata} shows that the Itoyori implementation excels in this pattern, outperforming all other implementations.
This performance advantage is intrinsic to the PGAS architecture.
Unlike message-passing models that require discrete serialization for each dependency, the Itoyori PGAS API supports aggregation of remote memory accesses via contiguous offset/length parameters.
This enables the runtime to fetch data for multiple dependencies in a single RDMA operation, drastically reducing the message injection rate.
The MPI and HPX implementations, constrained to sending individual messages for each dependency, suffer from substantial overheads.
The ItoyoriFBC implementation also underperforms relative to Itoyori, as its current implementation requires re-wrapping futures for each dependency (see~\cite{MiaItoyoriFBCAMTE24} for details), preventing contiguous memory optimization.
This finding emphasizes the potential of global address space abstractions for mitigating dense communication pattern overheads.
Experiments beyond 8 nodes with a width of 16 per core failed due to out-of-memory constraints imposed by the dense dependency structure.

\input{04fig4}
\input{04fig5}

\subsection{Discussion}
\textbf{The MPI implementation} remains the performance baseline for regular, static workloads.
Its ability to execute fine-grained tasks with minimal runtime overhead makes it optimal for applications where workload distribution is deterministic and known \textit{a priori}.
However, this performance comes at the expense of productivity.
As shown in Table~\ref{tab:program-metrics}, the MPI implementation requires greater development effort (137 LOC) compared to the Itoyori implementation (77 LOC).
This difference stems from explicit, low-level communication management in MPI versus Itoyori's abstract checkout/checkin API.

\textbf{The Itoyori implementation} achieves the highest productivity and outperforms MPI in communication-intensive scenarios (\emph{e.g.}, \texttt{all\_to\_all}) due to its PGAS cache.
However, its RDWS introduces noticeable overheads at low task granularities (METG), requiring sufficient parallelism (width) to amortize the RDWS costs.
This confirms that AMTs offer a viable alternative for irregular or communication-heavy applications where the development cost of explicit message passing is prohibitive.
While we do not isolate the checkout/checkin overhead separately, its cost is implicitly captured in the METG metric: the higher METG of Itoyori relative to MPI at low task widths reflects the combined overhead of RDWS and PGAS cache management, including checkout/checkin.
Isolating these individual overhead factors remains future work.

\textbf{The ItoyoriFBC implementation} enables the expression of dependencies as DAGs, theoretically offering superior handling of irregular, non-nested workloads.
However, in the controlled bulk-synchronous context of this study, this flexibility comes with a measurable overhead.
The requirement to explicitly manage future objects leads to a higher code complexity (115 LOC) and lower efficiency compared to the Itoyori implementation.
Furthermore, current limitations of ItoyoriFBC in handling shared dependencies, which require re-wrapping of futures, underscore the need for further optimization.

\textbf{The HPX implementation}, on the other hand, demonstrates superior performance in the load imbalance benchmarks.
However, its performance is sensitive to communication topology, with significant degradation observed in scenarios that concentrate traffic, such as the centralized pattern in the \texttt{spread} benchmark.
While its productivity metrics (224~LOC, 23~NLC) appear to lag behind those of MPI, this largely reflects the effort of adapting the MPI-centric Task Bench structure to HPX.
This observation suggests that while HPX is powerful for dynamic applications, it requires careful attention to communication patterns to fully realize its performance potential.

\textbf{In summary}:
For static, compute-bound problems, the MPI implementation remains the most efficient solution.
For dynamic, irregular applications requiring robust load balancing, HPX offers mature AMT programming.
For applications with complex communication patterns or those that benefit from global address space abstractions, the Itoyori implementation presents a superior combination of performance and productivity, provided that task granularity is sufficient to mask scheduling overheads.

%% file: 04fig1.tex
\begin{figure}[t]
    \centering
    \begin{subfigure}[t]{0.49\linewidth}
        \resizebox{0.99\linewidth}{!}{\input{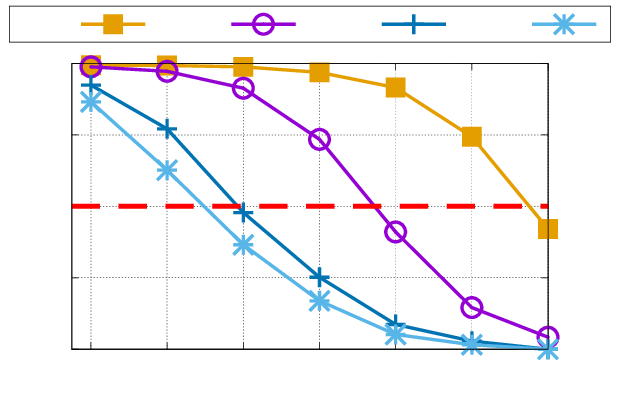}}
        \caption{1 width per core}
        \label{subfig:metgW1}
    \end{subfigure}\hfill
    \begin{subfigure}[t]{0.49\linewidth}
        \resizebox{0.99\linewidth}{!}{\input{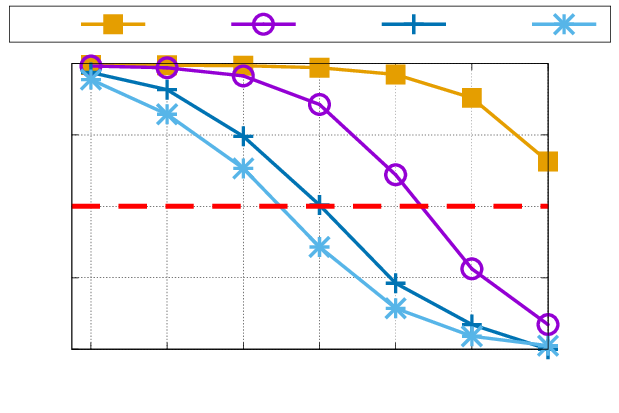}}
        \caption{4 width per core}
        \label{subfig:metgW4}
    \end{subfigure}

    \vspace*{5pt}

    \begin{subfigure}[t]{0.49\linewidth}
        \resizebox{0.99\linewidth}{!}{\input{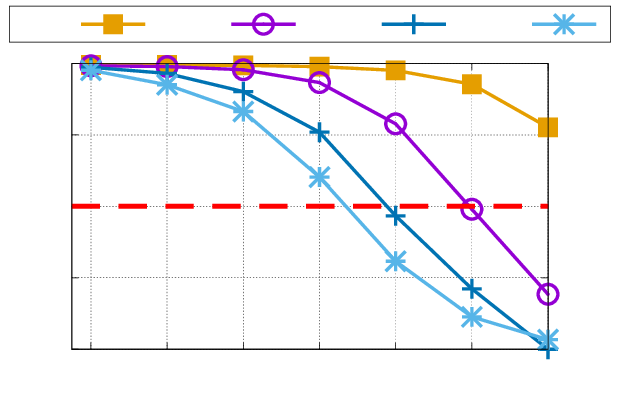}}
        \caption{16 width per core}
        \label{subfig:metgW16}
    \end{subfigure}\hfill
    \begin{subfigure}[t]{0.49\linewidth}
        \resizebox{0.99\linewidth}{!}{\input{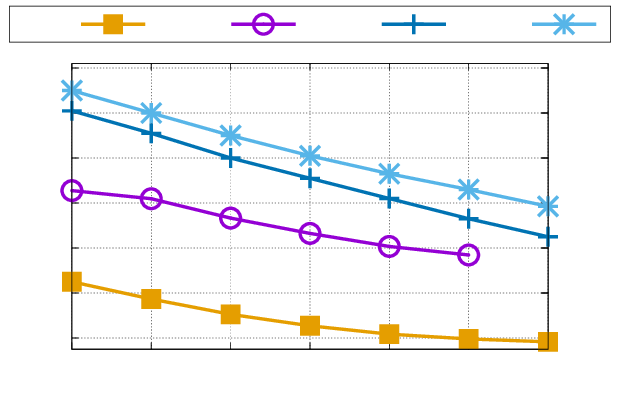}}
        \caption{METG~(50\%)}
        \label{subfig:metgN1}
    \end{subfigure}

    \caption{Efficiency and METG for varying widths with a \texttt{stencil} graph, \texttt{compute\_bound} kernel, and one node}
    \label{fig:varying-width}
\end{figure}

%% file: figs/compW1.tex
\begingroup
\Large
  \makeatletter
  \providecommand\color[2][]{%
    \GenericError{(gnuplot) \space\space\space\@spaces}{%
      Package color not loaded in conjunction with
      terminal option `colourtext'%
    }{See the gnuplot documentation for explanation.%
    }{Either use 'blacktext' in gnuplot or load the package
      color.sty in LaTeX.}%
    \renewcommand\color[2][]{}%
  }%
  \providecommand\includegraphics[2][]{%
    \GenericError{(gnuplot) \space\space\space\@spaces}{%
      Package graphicx or graphics not loaded%
    }{See the gnuplot documentation for explanation.%
    }{The gnuplot epslatex terminal needs graphicx.sty or graphics.sty.}%
    \renewcommand\includegraphics[2][]{}%
  }%
  \providecommand\rotatebox[2]{#2}%
  \@ifundefined{ifGPcolor}{%
    \newif\ifGPcolor
    \GPcolortrue
  }{}%
  \@ifundefined{ifGPblacktext}{%
    \newif\ifGPblacktext
    \GPblacktextfalse
  }{}%
  \let\gplgaddtomacro\g@addto@macro
  \gdef\gplbacktext{}%
  \gdef\gplfronttext{}%
  \makeatother
  \ifGPblacktext
    \def\colorrgb#1{}%
    \def\colorgray#1{}%
  \else
    \ifGPcolor
      \def\colorrgb#1{\color[rgb]{#1}}%
      \def\colorgray#1{\color[gray]{#1}}%
      \expandafter\def\csname LTw\endcsname{\color{white}}%
      \expandafter\def\csname LTb\endcsname{\color{black}}%
      \expandafter\def\csname LTa\endcsname{\color{black}}%
      \expandafter\def\csname LT0\endcsname{\color[rgb]{1,0,0}}%
      \expandafter\def\csname LT1\endcsname{\color[rgb]{0,1,0}}%
      \expandafter\def\csname LT2\endcsname{\color[rgb]{0,0,1}}%
      \expandafter\def\csname LT3\endcsname{\color[rgb]{1,0,1}}%
      \expandafter\def\csname LT4\endcsname{\color[rgb]{0,1,1}}%
      \expandafter\def\csname LT5\endcsname{\color[rgb]{1,1,0}}%
      \expandafter\def\csname LT6\endcsname{\color[rgb]{0,0,0}}%
      \expandafter\def\csname LT7\endcsname{\color[rgb]{1,0.3,0}}%
      \expandafter\def\csname LT8\endcsname{\color[rgb]{0.5,0.5,0.5}}%
    \else
      \def\colorrgb#1{\color{black}}%
      \def\colorgray#1{\color[gray]{#1}}%
      \expandafter\def\csname LTw\endcsname{\color{white}}%
      \expandafter\def\csname LTb\endcsname{\color{black}}%
      \expandafter\def\csname LTa\endcsname{\color{black}}%
      \expandafter\def\csname LT0\endcsname{\color{black}}%
      \expandafter\def\csname LT1\endcsname{\color{black}}%
      \expandafter\def\csname LT2\endcsname{\color{black}}%
      \expandafter\def\csname LT3\endcsname{\color{black}}%
      \expandafter\def\csname LT4\endcsname{\color{black}}%
      \expandafter\def\csname LT5\endcsname{\color{black}}%
      \expandafter\def\csname LT6\endcsname{\color{black}}%
      \expandafter\def\csname LT7\endcsname{\color{black}}%
      \expandafter\def\csname LT8\endcsname{\color{black}}%
    \fi
  \fi
    \setlength{\unitlength}{0.0500bp}%
    \ifx\gptboxheight\undefined%
      \newlength{\gptboxheight}%
      \newlength{\gptboxwidth}%
      \newsavebox{\gptboxtext}%
    \fi%
    \setlength{\fboxrule}{0.5pt}%
    \setlength{\fboxsep}{1pt}%
    \definecolor{tbcol}{rgb}{1,1,1}%
\begin{picture}(5952.00,3968.00)%
    \gplgaddtomacro\gplbacktext{%
      \csname LTb\endcsname
      \put(552,612){\makebox(0,0)[r]{\strut{}0\%}}%
      \csname LTb\endcsname
      \put(552,1298){\makebox(0,0)[r]{\strut{}25\%}}%
      \csname LTb\endcsname
      \put(552,1984){\makebox(0,0)[r]{\strut{}50\%}}%
      \csname LTb\endcsname
      \put(552,2669){\makebox(0,0)[r]{\strut{}75\%}}%
      \csname LTb\endcsname
      \put(552,3355){\makebox(0,0)[r]{\strut{}100\%}}%
      \csname LTb\endcsname
      \put(5261,382){\makebox(0,0){\strut{}$2^{8}$}}%
      \csname LTb\endcsname
      \put(4530,382){\makebox(0,0){\strut{}$2^{10}$}}%
      \csname LTb\endcsname
      \put(3798,382){\makebox(0,0){\strut{}$2^{12}$}}%
      \csname LTb\endcsname
      \put(3067,382){\makebox(0,0){\strut{}$2^{14}$}}%
      \csname LTb\endcsname
      \put(2336,382){\makebox(0,0){\strut{}$2^{16}$}}%
      \csname LTb\endcsname
      \put(1604,382){\makebox(0,0){\strut{}$2^{18}$}}%
      \csname LTb\endcsname
      \put(873,382){\makebox(0,0){\strut{}$2^{20}$}}%
    }%
    \gplgaddtomacro\gplfronttext{%
      \csname LTb\endcsname
      \put(641,3732){\makebox(0,0)[r]{\strut{}MPI}}%
      \csname LTb\endcsname
      \put(2084,3732){\makebox(0,0)[r]{\strut{}HPX}}%
      \csname LTb\endcsname
      \put(3527,3732){\makebox(0,0)[r]{\strut{}ITO}}%
      \csname LTb\endcsname
      \put(4970,3732){\makebox(0,0)[r]{\strut{}FBC}}%
      \csname LTb\endcsname
      \put(-356,1983){\rotatebox{-270.00}{\makebox(0,0){\strut{}Efficiency}}}%
      \put(2975,37){\makebox(0,0){\strut{}Task-Size}}%
    }%
    \gplbacktext
    \put(0,0){\includegraphics[width={297.60bp},height={198.40bp}]{compW1}}%
    \gplfronttext
  \end{picture}%
\endgroup

%% file: figs/compW4.tex
\begingroup
\Large
  \makeatletter
  \providecommand\color[2][]{%
    \GenericError{(gnuplot) \space\space\space\@spaces}{%
      Package color not loaded in conjunction with
      terminal option `colourtext'%
    }{See the gnuplot documentation for explanation.%
    }{Either use 'blacktext' in gnuplot or load the package
      color.sty in LaTeX.}%
    \renewcommand\color[2][]{}%
  }%
  \providecommand\includegraphics[2][]{%
    \GenericError{(gnuplot) \space\space\space\@spaces}{%
      Package graphicx or graphics not loaded%
    }{See the gnuplot documentation for explanation.%
    }{The gnuplot epslatex terminal needs graphicx.sty or graphics.sty.}%
    \renewcommand\includegraphics[2][]{}%
  }%
  \providecommand\rotatebox[2]{#2}%
  \@ifundefined{ifGPcolor}{%
    \newif\ifGPcolor
    \GPcolortrue
  }{}%
  \@ifundefined{ifGPblacktext}{%
    \newif\ifGPblacktext
    \GPblacktextfalse
  }{}%
  \let\gplgaddtomacro\g@addto@macro
  \gdef\gplbacktext{}%
  \gdef\gplfronttext{}%
  \makeatother
  \ifGPblacktext
    \def\colorrgb#1{}%
    \def\colorgray#1{}%
  \else
    \ifGPcolor
      \def\colorrgb#1{\color[rgb]{#1}}%
      \def\colorgray#1{\color[gray]{#1}}%
      \expandafter\def\csname LTw\endcsname{\color{white}}%
      \expandafter\def\csname LTb\endcsname{\color{black}}%
      \expandafter\def\csname LTa\endcsname{\color{black}}%
      \expandafter\def\csname LT0\endcsname{\color[rgb]{1,0,0}}%
      \expandafter\def\csname LT1\endcsname{\color[rgb]{0,1,0}}%
      \expandafter\def\csname LT2\endcsname{\color[rgb]{0,0,1}}%
      \expandafter\def\csname LT3\endcsname{\color[rgb]{1,0,1}}%
      \expandafter\def\csname LT4\endcsname{\color[rgb]{0,1,1}}%
      \expandafter\def\csname LT5\endcsname{\color[rgb]{1,1,0}}%
      \expandafter\def\csname LT6\endcsname{\color[rgb]{0,0,0}}%
      \expandafter\def\csname LT7\endcsname{\color[rgb]{1,0.3,0}}%
      \expandafter\def\csname LT8\endcsname{\color[rgb]{0.5,0.5,0.5}}%
    \else
      \def\colorrgb#1{\color{black}}%
      \def\colorgray#1{\color[gray]{#1}}%
      \expandafter\def\csname LTw\endcsname{\color{white}}%
      \expandafter\def\csname LTb\endcsname{\color{black}}%
      \expandafter\def\csname LTa\endcsname{\color{black}}%
      \expandafter\def\csname LT0\endcsname{\color{black}}%
      \expandafter\def\csname LT1\endcsname{\color{black}}%
      \expandafter\def\csname LT2\endcsname{\color{black}}%
      \expandafter\def\csname LT3\endcsname{\color{black}}%
      \expandafter\def\csname LT4\endcsname{\color{black}}%
      \expandafter\def\csname LT5\endcsname{\color{black}}%
      \expandafter\def\csname LT6\endcsname{\color{black}}%
      \expandafter\def\csname LT7\endcsname{\color{black}}%
      \expandafter\def\csname LT8\endcsname{\color{black}}%
    \fi
  \fi
    \setlength{\unitlength}{0.0500bp}%
    \ifx\gptboxheight\undefined%
      \newlength{\gptboxheight}%
      \newlength{\gptboxwidth}%
      \newsavebox{\gptboxtext}%
    \fi%
    \setlength{\fboxrule}{0.5pt}%
    \setlength{\fboxsep}{1pt}%
    \definecolor{tbcol}{rgb}{1,1,1}%
\begin{picture}(5952.00,3968.00)%
    \gplgaddtomacro\gplbacktext{%
      \csname LTb\endcsname
      \put(552,612){\makebox(0,0)[r]{\strut{}0\%}}%
      \csname LTb\endcsname
      \put(552,1298){\makebox(0,0)[r]{\strut{}25\%}}%
      \csname LTb\endcsname
      \put(552,1984){\makebox(0,0)[r]{\strut{}50\%}}%
      \csname LTb\endcsname
      \put(552,2669){\makebox(0,0)[r]{\strut{}75\%}}%
      \csname LTb\endcsname
      \put(552,3355){\makebox(0,0)[r]{\strut{}100\%}}%
      \csname LTb\endcsname
      \put(5261,382){\makebox(0,0){\strut{}$2^{8}$}}%
      \csname LTb\endcsname
      \put(4530,382){\makebox(0,0){\strut{}$2^{10}$}}%
      \csname LTb\endcsname
      \put(3798,382){\makebox(0,0){\strut{}$2^{12}$}}%
      \csname LTb\endcsname
      \put(3067,382){\makebox(0,0){\strut{}$2^{14}$}}%
      \csname LTb\endcsname
      \put(2336,382){\makebox(0,0){\strut{}$2^{16}$}}%
      \csname LTb\endcsname
      \put(1604,382){\makebox(0,0){\strut{}$2^{18}$}}%
      \csname LTb\endcsname
      \put(873,382){\makebox(0,0){\strut{}$2^{20}$}}%
    }%
    \gplgaddtomacro\gplfronttext{%
      \csname LTb\endcsname
      \put(641,3732){\makebox(0,0)[r]{\strut{}MPI}}%
      \csname LTb\endcsname
      \put(2084,3732){\makebox(0,0)[r]{\strut{}HPX}}%
      \csname LTb\endcsname
      \put(3527,3732){\makebox(0,0)[r]{\strut{}ITO}}%
      \csname LTb\endcsname
      \put(4970,3732){\makebox(0,0)[r]{\strut{}FBC}}%
      \csname LTb\endcsname
      \put(-356,1983){\rotatebox{-270.00}{\makebox(0,0){\strut{}Efficiency}}}%
      \put(2975,37){\makebox(0,0){\strut{}Task-Size}}%
    }%
    \gplbacktext
    \put(0,0){\includegraphics[width={297.60bp},height={198.40bp}]{compW4}}%
    \gplfronttext
  \end{picture}%
\endgroup

%% file: figs/compW16.tex
\begingroup
\Large
  \makeatletter
  \providecommand\color[2][]{%
    \GenericError{(gnuplot) \space\space\space\@spaces}{%
      Package color not loaded in conjunction with
      terminal option `colourtext'%
    }{See the gnuplot documentation for explanation.%
    }{Either use 'blacktext' in gnuplot or load the package
      color.sty in LaTeX.}%
    \renewcommand\color[2][]{}%
  }%
  \providecommand\includegraphics[2][]{%
    \GenericError{(gnuplot) \space\space\space\@spaces}{%
      Package graphicx or graphics not loaded%
    }{See the gnuplot documentation for explanation.%
    }{The gnuplot epslatex terminal needs graphicx.sty or graphics.sty.}%
    \renewcommand\includegraphics[2][]{}%
  }%
  \providecommand\rotatebox[2]{#2}%
  \@ifundefined{ifGPcolor}{%
    \newif\ifGPcolor
    \GPcolortrue
  }{}%
  \@ifundefined{ifGPblacktext}{%
    \newif\ifGPblacktext
    \GPblacktextfalse
  }{}%
  \let\gplgaddtomacro\g@addto@macro
  \gdef\gplbacktext{}%
  \gdef\gplfronttext{}%
  \makeatother
  \ifGPblacktext
    \def\colorrgb#1{}%
    \def\colorgray#1{}%
  \else
    \ifGPcolor
      \def\colorrgb#1{\color[rgb]{#1}}%
      \def\colorgray#1{\color[gray]{#1}}%
      \expandafter\def\csname LTw\endcsname{\color{white}}%
      \expandafter\def\csname LTb\endcsname{\color{black}}%
      \expandafter\def\csname LTa\endcsname{\color{black}}%
      \expandafter\def\csname LT0\endcsname{\color[rgb]{1,0,0}}%
      \expandafter\def\csname LT1\endcsname{\color[rgb]{0,1,0}}%
      \expandafter\def\csname LT2\endcsname{\color[rgb]{0,0,1}}%
      \expandafter\def\csname LT3\endcsname{\color[rgb]{1,0,1}}%
      \expandafter\def\csname LT4\endcsname{\color[rgb]{0,1,1}}%
      \expandafter\def\csname LT5\endcsname{\color[rgb]{1,1,0}}%
      \expandafter\def\csname LT6\endcsname{\color[rgb]{0,0,0}}%
      \expandafter\def\csname LT7\endcsname{\color[rgb]{1,0.3,0}}%
      \expandafter\def\csname LT8\endcsname{\color[rgb]{0.5,0.5,0.5}}%
    \else
      \def\colorrgb#1{\color{black}}%
      \def\colorgray#1{\color[gray]{#1}}%
      \expandafter\def\csname LTw\endcsname{\color{white}}%
      \expandafter\def\csname LTb\endcsname{\color{black}}%
      \expandafter\def\csname LTa\endcsname{\color{black}}%
      \expandafter\def\csname LT0\endcsname{\color{black}}%
      \expandafter\def\csname LT1\endcsname{\color{black}}%
      \expandafter\def\csname LT2\endcsname{\color{black}}%
      \expandafter\def\csname LT3\endcsname{\color{black}}%
      \expandafter\def\csname LT4\endcsname{\color{black}}%
      \expandafter\def\csname LT5\endcsname{\color{black}}%
      \expandafter\def\csname LT6\endcsname{\color{black}}%
      \expandafter\def\csname LT7\endcsname{\color{black}}%
      \expandafter\def\csname LT8\endcsname{\color{black}}%
    \fi
  \fi
    \setlength{\unitlength}{0.0500bp}%
    \ifx\gptboxheight\undefined%
      \newlength{\gptboxheight}%
      \newlength{\gptboxwidth}%
      \newsavebox{\gptboxtext}%
    \fi%
    \setlength{\fboxrule}{0.5pt}%
    \setlength{\fboxsep}{1pt}%
    \definecolor{tbcol}{rgb}{1,1,1}%
\begin{picture}(5952.00,3968.00)%
    \gplgaddtomacro\gplbacktext{%
      \csname LTb\endcsname
      \put(552,612){\makebox(0,0)[r]{\strut{}0\%}}%
      \csname LTb\endcsname
      \put(552,1298){\makebox(0,0)[r]{\strut{}25\%}}%
      \csname LTb\endcsname
      \put(552,1984){\makebox(0,0)[r]{\strut{}50\%}}%
      \csname LTb\endcsname
      \put(552,2669){\makebox(0,0)[r]{\strut{}75\%}}%
      \csname LTb\endcsname
      \put(552,3355){\makebox(0,0)[r]{\strut{}100\%}}%
      \csname LTb\endcsname
      \put(5261,382){\makebox(0,0){\strut{}$2^{8}$}}%
      \csname LTb\endcsname
      \put(4530,382){\makebox(0,0){\strut{}$2^{10}$}}%
      \csname LTb\endcsname
      \put(3798,382){\makebox(0,0){\strut{}$2^{12}$}}%
      \csname LTb\endcsname
      \put(3067,382){\makebox(0,0){\strut{}$2^{14}$}}%
      \csname LTb\endcsname
      \put(2336,382){\makebox(0,0){\strut{}$2^{16}$}}%
      \csname LTb\endcsname
      \put(1604,382){\makebox(0,0){\strut{}$2^{18}$}}%
      \csname LTb\endcsname
      \put(873,382){\makebox(0,0){\strut{}$2^{20}$}}%
    }%
    \gplgaddtomacro\gplfronttext{%
      \csname LTb\endcsname
      \put(641,3732){\makebox(0,0)[r]{\strut{}MPI}}%
      \csname LTb\endcsname
      \put(2084,3732){\makebox(0,0)[r]{\strut{}HPX}}%
      \csname LTb\endcsname
      \put(3527,3732){\makebox(0,0)[r]{\strut{}ITO}}%
      \csname LTb\endcsname
      \put(4970,3732){\makebox(0,0)[r]{\strut{}FBC}}%
      \csname LTb\endcsname
      \put(-356,1983){\rotatebox{-270.00}{\makebox(0,0){\strut{}Efficiency}}}%
      \put(2975,37){\makebox(0,0){\strut{}Task-Size}}%
    }%
    \gplbacktext
    \put(0,0){\includegraphics[width={297.60bp},height={198.40bp}]{compW16}}%
    \gplfronttext
  \end{picture}%
\endgroup

%% file: figs/metgN1.tex
\begingroup
\Large
  \makeatletter
  \providecommand\color[2][]{%
    \GenericError{(gnuplot) \space\space\space\@spaces}{%
      Package color not loaded in conjunction with
      terminal option `colourtext'%
    }{See the gnuplot documentation for explanation.%
    }{Either use 'blacktext' in gnuplot or load the package
      color.sty in LaTeX.}%
    \renewcommand\color[2][]{}%
  }%
  \providecommand\includegraphics[2][]{%
    \GenericError{(gnuplot) \space\space\space\@spaces}{%
      Package graphicx or graphics not loaded%
    }{See the gnuplot documentation for explanation.%
    }{The gnuplot epslatex terminal needs graphicx.sty or graphics.sty.}%
    \renewcommand\includegraphics[2][]{}%
  }%
  \providecommand\rotatebox[2]{#2}%
  \@ifundefined{ifGPcolor}{%
    \newif\ifGPcolor
    \GPcolortrue
  }{}%
  \@ifundefined{ifGPblacktext}{%
    \newif\ifGPblacktext
    \GPblacktextfalse
  }{}%
  \let\gplgaddtomacro\g@addto@macro
  \gdef\gplbacktext{}%
  \gdef\gplfronttext{}%
  \makeatother
  \ifGPblacktext
    \def\colorrgb#1{}%
    \def\colorgray#1{}%
  \else
    \ifGPcolor
      \def\colorrgb#1{\color[rgb]{#1}}%
      \def\colorgray#1{\color[gray]{#1}}%
      \expandafter\def\csname LTw\endcsname{\color{white}}%
      \expandafter\def\csname LTb\endcsname{\color{black}}%
      \expandafter\def\csname LTa\endcsname{\color{black}}%
      \expandafter\def\csname LT0\endcsname{\color[rgb]{1,0,0}}%
      \expandafter\def\csname LT1\endcsname{\color[rgb]{0,1,0}}%
      \expandafter\def\csname LT2\endcsname{\color[rgb]{0,0,1}}%
      \expandafter\def\csname LT3\endcsname{\color[rgb]{1,0,1}}%
      \expandafter\def\csname LT4\endcsname{\color[rgb]{0,1,1}}%
      \expandafter\def\csname LT5\endcsname{\color[rgb]{1,1,0}}%
      \expandafter\def\csname LT6\endcsname{\color[rgb]{0,0,0}}%
      \expandafter\def\csname LT7\endcsname{\color[rgb]{1,0.3,0}}%
      \expandafter\def\csname LT8\endcsname{\color[rgb]{0.5,0.5,0.5}}%
    \else
      \def\colorrgb#1{\color{black}}%
      \def\colorgray#1{\color[gray]{#1}}%
      \expandafter\def\csname LTw\endcsname{\color{white}}%
      \expandafter\def\csname LTb\endcsname{\color{black}}%
      \expandafter\def\csname LTa\endcsname{\color{black}}%
      \expandafter\def\csname LT0\endcsname{\color{black}}%
      \expandafter\def\csname LT1\endcsname{\color{black}}%
      \expandafter\def\csname LT2\endcsname{\color{black}}%
      \expandafter\def\csname LT3\endcsname{\color{black}}%
      \expandafter\def\csname LT4\endcsname{\color{black}}%
      \expandafter\def\csname LT5\endcsname{\color{black}}%
      \expandafter\def\csname LT6\endcsname{\color{black}}%
      \expandafter\def\csname LT7\endcsname{\color{black}}%
      \expandafter\def\csname LT8\endcsname{\color{black}}%
    \fi
  \fi
    \setlength{\unitlength}{0.0500bp}%
    \ifx\gptboxheight\undefined%
      \newlength{\gptboxheight}%
      \newlength{\gptboxwidth}%
      \newsavebox{\gptboxtext}%
    \fi%
    \setlength{\fboxrule}{0.5pt}%
    \setlength{\fboxsep}{1pt}%
    \definecolor{tbcol}{rgb}{1,1,1}%
\begin{picture}(5952.00,3968.00)%
    \gplgaddtomacro\gplbacktext{%
      \csname LTb\endcsname
      \put(552,720){\makebox(0,0)[r]{\strut{}$2^{6}$}}%
      \csname LTb\endcsname
      \put(552,1152){\makebox(0,0)[r]{\strut{}$2^{8}$}}%
      \csname LTb\endcsname
      \put(552,1584){\makebox(0,0)[r]{\strut{}$2^{10}$}}%
      \csname LTb\endcsname
      \put(552,2016){\makebox(0,0)[r]{\strut{}$2^{12}$}}%
      \csname LTb\endcsname
      \put(552,2448){\makebox(0,0)[r]{\strut{}$2^{14}$}}%
      \csname LTb\endcsname
      \put(552,2880){\makebox(0,0)[r]{\strut{}$2^{16}$}}%
      \csname LTb\endcsname
      \put(552,3312){\makebox(0,0)[r]{\strut{}$2^{18}$}}%
      \csname LTb\endcsname
      \put(690,382){\makebox(0,0){\strut{}$1$}}%
      \csname LTb\endcsname
      \put(1452,382){\makebox(0,0){\strut{}$2$}}%
      \csname LTb\endcsname
      \put(2214,382){\makebox(0,0){\strut{}$4$}}%
      \csname LTb\endcsname
      \put(2976,382){\makebox(0,0){\strut{}$8$}}%
      \csname LTb\endcsname
      \put(3737,382){\makebox(0,0){\strut{}$16$}}%
      \csname LTb\endcsname
      \put(4499,382){\makebox(0,0){\strut{}$32$}}%
      \csname LTb\endcsname
      \put(5261,382){\makebox(0,0){\strut{}$64$}}%
      \put(5399,720){\makebox(0,0)[l]{\strut{}$0.19\mu s$}}%
      \put(5399,1152){\makebox(0,0)[l]{\strut{}$0.79\mu s$}}%
      \put(5399,1584){\makebox(0,0)[l]{\strut{}$3.16\mu s$}}%
      \put(5399,2016){\makebox(0,0)[l]{\strut{}$12.6\mu s$}}%
      \put(5399,2448){\makebox(0,0)[l]{\strut{}$50.5\mu s$}}%
      \put(5399,2880){\makebox(0,0)[l]{\strut{}$202\mu s$}}%
      \put(5399,3312){\makebox(0,0)[l]{\strut{}$808\mu s$}}%
    }%
    \gplgaddtomacro\gplfronttext{%
      \csname LTb\endcsname
      \put(641,3732){\makebox(0,0)[r]{\strut{}MPI}}%
      \csname LTb\endcsname
      \put(2084,3732){\makebox(0,0)[r]{\strut{}HPX}}%
      \csname LTb\endcsname
      \put(3527,3732){\makebox(0,0)[r]{\strut{}ITO}}%
      \csname LTb\endcsname
      \put(4970,3732){\makebox(0,0)[r]{\strut{}FBC}}%
      \csname LTb\endcsname
      \put(-80,1983){\rotatebox{-270.00}{\makebox(0,0){\strut{}Kernel Iterations}}}%
      \put(6618,1983){\rotatebox{-270.00}{\makebox(0,0){\strut{}Kernel Runtime}}}%
      \put(2975,37){\makebox(0,0){\strut{}Width per core}}%
    }%
    \gplbacktext
    \put(0,0){\includegraphics[width={297.60bp},height={198.40bp}]{metgN1}}%
    \gplfronttext
  \end{picture}%
\endgroup

%% file: 04fig2.tex
\begin{figure}[t]
    \centering
    \begin{subfigure}[t]{0.49\linewidth}
        \resizebox{0.99\linewidth}{!}{\input{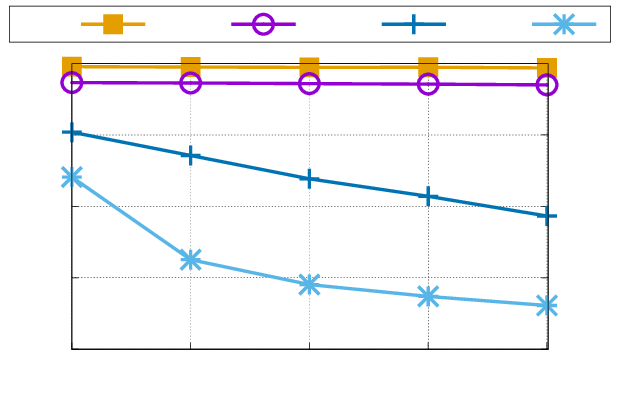}}
        \caption{Weak scaling with task size $2^{14}$}
        \label{subfig:metgW16Scaling}
    \end{subfigure}\hfill
    \begin{subfigure}[t]{0.49\linewidth}
        \resizebox{0.99\linewidth}{!}{\input{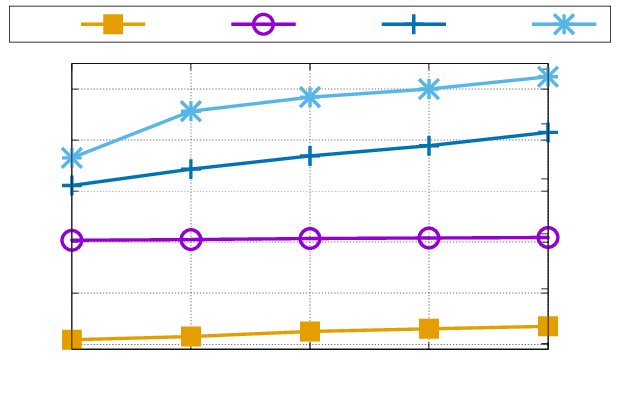}}
        \caption{METG~(50\%) for weak scaling}
        \label{subfig:metgWeakScaled}
    \end{subfigure}

    \caption{Weak scaling from 1 to 16 nodes for a \texttt{stencil} graph, \texttt{compute\_bound} kernel, and 16 width per core}
    \label{fig:weak-scaling}
\end{figure}

%% file: figs/compSize14WeakScale.tex
\begingroup
\Large
  \makeatletter
  \providecommand\color[2][]{%
    \GenericError{(gnuplot) \space\space\space\@spaces}{%
      Package color not loaded in conjunction with
      terminal option `colourtext'%
    }{See the gnuplot documentation for explanation.%
    }{Either use 'blacktext' in gnuplot or load the package
      color.sty in LaTeX.}%
    \renewcommand\color[2][]{}%
  }%
  \providecommand\includegraphics[2][]{%
    \GenericError{(gnuplot) \space\space\space\@spaces}{%
      Package graphicx or graphics not loaded%
    }{See the gnuplot documentation for explanation.%
    }{The gnuplot epslatex terminal needs graphicx.sty or graphics.sty.}%
    \renewcommand\includegraphics[2][]{}%
  }%
  \providecommand\rotatebox[2]{#2}%
  \@ifundefined{ifGPcolor}{%
    \newif\ifGPcolor
    \GPcolortrue
  }{}%
  \@ifundefined{ifGPblacktext}{%
    \newif\ifGPblacktext
    \GPblacktextfalse
  }{}%
  \let\gplgaddtomacro\g@addto@macro
  \gdef\gplbacktext{}%
  \gdef\gplfronttext{}%
  \makeatother
  \ifGPblacktext
    \def\colorrgb#1{}%
    \def\colorgray#1{}%
  \else
    \ifGPcolor
      \def\colorrgb#1{\color[rgb]{#1}}%
      \def\colorgray#1{\color[gray]{#1}}%
      \expandafter\def\csname LTw\endcsname{\color{white}}%
      \expandafter\def\csname LTb\endcsname{\color{black}}%
      \expandafter\def\csname LTa\endcsname{\color{black}}%
      \expandafter\def\csname LT0\endcsname{\color[rgb]{1,0,0}}%
      \expandafter\def\csname LT1\endcsname{\color[rgb]{0,1,0}}%
      \expandafter\def\csname LT2\endcsname{\color[rgb]{0,0,1}}%
      \expandafter\def\csname LT3\endcsname{\color[rgb]{1,0,1}}%
      \expandafter\def\csname LT4\endcsname{\color[rgb]{0,1,1}}%
      \expandafter\def\csname LT5\endcsname{\color[rgb]{1,1,0}}%
      \expandafter\def\csname LT6\endcsname{\color[rgb]{0,0,0}}%
      \expandafter\def\csname LT7\endcsname{\color[rgb]{1,0.3,0}}%
      \expandafter\def\csname LT8\endcsname{\color[rgb]{0.5,0.5,0.5}}%
    \else
      \def\colorrgb#1{\color{black}}%
      \def\colorgray#1{\color[gray]{#1}}%
      \expandafter\def\csname LTw\endcsname{\color{white}}%
      \expandafter\def\csname LTb\endcsname{\color{black}}%
      \expandafter\def\csname LTa\endcsname{\color{black}}%
      \expandafter\def\csname LT0\endcsname{\color{black}}%
      \expandafter\def\csname LT1\endcsname{\color{black}}%
      \expandafter\def\csname LT2\endcsname{\color{black}}%
      \expandafter\def\csname LT3\endcsname{\color{black}}%
      \expandafter\def\csname LT4\endcsname{\color{black}}%
      \expandafter\def\csname LT5\endcsname{\color{black}}%
      \expandafter\def\csname LT6\endcsname{\color{black}}%
      \expandafter\def\csname LT7\endcsname{\color{black}}%
      \expandafter\def\csname LT8\endcsname{\color{black}}%
    \fi
  \fi
    \setlength{\unitlength}{0.0500bp}%
    \ifx\gptboxheight\undefined%
      \newlength{\gptboxheight}%
      \newlength{\gptboxwidth}%
      \newsavebox{\gptboxtext}%
    \fi%
    \setlength{\fboxrule}{0.5pt}%
    \setlength{\fboxsep}{1pt}%
    \definecolor{tbcol}{rgb}{1,1,1}%
\begin{picture}(5952.00,3968.00)%
    \gplgaddtomacro\gplbacktext{%
      \csname LTb\endcsname
      \put(552,612){\makebox(0,0)[r]{\strut{}0\%}}%
      \csname LTb\endcsname
      \put(552,1298){\makebox(0,0)[r]{\strut{}25\%}}%
      \csname LTb\endcsname
      \put(552,1984){\makebox(0,0)[r]{\strut{}50\%}}%
      \csname LTb\endcsname
      \put(552,2669){\makebox(0,0)[r]{\strut{}75\%}}%
      \csname LTb\endcsname
      \put(552,3355){\makebox(0,0)[r]{\strut{}100\%}}%
      \csname LTb\endcsname
      \put(690,382){\makebox(0,0){\strut{}$1$}}%
      \csname LTb\endcsname
      \put(1830,382){\makebox(0,0){\strut{}$2$}}%
      \csname LTb\endcsname
      \put(2970,382){\makebox(0,0){\strut{}$4$}}%
      \csname LTb\endcsname
      \put(4111,382){\makebox(0,0){\strut{}$8$}}%
      \csname LTb\endcsname
      \put(5251,382){\makebox(0,0){\strut{}$16$}}%
    }%
    \gplgaddtomacro\gplfronttext{%
      \csname LTb\endcsname
      \put(641,3732){\makebox(0,0)[r]{\strut{}MPI}}%
      \csname LTb\endcsname
      \put(2084,3732){\makebox(0,0)[r]{\strut{}HPX}}%
      \csname LTb\endcsname
      \put(3527,3732){\makebox(0,0)[r]{\strut{}ITO}}%
      \csname LTb\endcsname
      \put(4970,3732){\makebox(0,0)[r]{\strut{}FBC}}%
      \csname LTb\endcsname
      \put(-356,1983){\rotatebox{-270.00}{\makebox(0,0){\strut{}Efficiency}}}%
      \put(2975,37){\makebox(0,0){\strut{}Number of Nodes}}%
    }%
    \gplbacktext
    \put(0,0){\includegraphics[width={297.60bp},height={198.40bp}]{compSize14WeakScale}}%
    \gplfronttext
  \end{picture}%
\endgroup

%% file: figs/metgWeakScaling.tex
\begingroup
\Large
  \makeatletter
  \providecommand\color[2][]{%
    \GenericError{(gnuplot) \space\space\space\@spaces}{%
      Package color not loaded in conjunction with
      terminal option `colourtext'%
    }{See the gnuplot documentation for explanation.%
    }{Either use 'blacktext' in gnuplot or load the package
      color.sty in LaTeX.}%
    \renewcommand\color[2][]{}%
  }%
  \providecommand\includegraphics[2][]{%
    \GenericError{(gnuplot) \space\space\space\@spaces}{%
      Package graphicx or graphics not loaded%
    }{See the gnuplot documentation for explanation.%
    }{The gnuplot epslatex terminal needs graphicx.sty or graphics.sty.}%
    \renewcommand\includegraphics[2][]{}%
  }%
  \providecommand\rotatebox[2]{#2}%
  \@ifundefined{ifGPcolor}{%
    \newif\ifGPcolor
    \GPcolortrue
  }{}%
  \@ifundefined{ifGPblacktext}{%
    \newif\ifGPblacktext
    \GPblacktextfalse
  }{}%
  \let\gplgaddtomacro\g@addto@macro
  \gdef\gplbacktext{}%
  \gdef\gplfronttext{}%
  \makeatother
  \ifGPblacktext
    \def\colorrgb#1{}%
    \def\colorgray#1{}%
  \else
    \ifGPcolor
      \def\colorrgb#1{\color[rgb]{#1}}%
      \def\colorgray#1{\color[gray]{#1}}%
      \expandafter\def\csname LTw\endcsname{\color{white}}%
      \expandafter\def\csname LTb\endcsname{\color{black}}%
      \expandafter\def\csname LTa\endcsname{\color{black}}%
      \expandafter\def\csname LT0\endcsname{\color[rgb]{1,0,0}}%
      \expandafter\def\csname LT1\endcsname{\color[rgb]{0,1,0}}%
      \expandafter\def\csname LT2\endcsname{\color[rgb]{0,0,1}}%
      \expandafter\def\csname LT3\endcsname{\color[rgb]{1,0,1}}%
      \expandafter\def\csname LT4\endcsname{\color[rgb]{0,1,1}}%
      \expandafter\def\csname LT5\endcsname{\color[rgb]{1,1,0}}%
      \expandafter\def\csname LT6\endcsname{\color[rgb]{0,0,0}}%
      \expandafter\def\csname LT7\endcsname{\color[rgb]{1,0.3,0}}%
      \expandafter\def\csname LT8\endcsname{\color[rgb]{0.5,0.5,0.5}}%
    \else
      \def\colorrgb#1{\color{black}}%
      \def\colorgray#1{\color[gray]{#1}}%
      \expandafter\def\csname LTw\endcsname{\color{white}}%
      \expandafter\def\csname LTb\endcsname{\color{black}}%
      \expandafter\def\csname LTa\endcsname{\color{black}}%
      \expandafter\def\csname LT0\endcsname{\color{black}}%
      \expandafter\def\csname LT1\endcsname{\color{black}}%
      \expandafter\def\csname LT2\endcsname{\color{black}}%
      \expandafter\def\csname LT3\endcsname{\color{black}}%
      \expandafter\def\csname LT4\endcsname{\color{black}}%
      \expandafter\def\csname LT5\endcsname{\color{black}}%
      \expandafter\def\csname LT6\endcsname{\color{black}}%
      \expandafter\def\csname LT7\endcsname{\color{black}}%
      \expandafter\def\csname LT8\endcsname{\color{black}}%
    \fi
  \fi
    \setlength{\unitlength}{0.0500bp}%
    \ifx\gptboxheight\undefined%
      \newlength{\gptboxheight}%
      \newlength{\gptboxwidth}%
      \newsavebox{\gptboxtext}%
    \fi%
    \setlength{\fboxrule}{0.5pt}%
    \setlength{\fboxsep}{1pt}%
    \definecolor{tbcol}{rgb}{1,1,1}%
\begin{picture}(5952.00,3968.00)%
    \gplgaddtomacro\gplbacktext{%
      \csname LTb\endcsname
      \put(552,661){\makebox(0,0)[r]{\strut{}$2^{6}$}}%
      \csname LTb\endcsname
      \put(552,1151){\makebox(0,0)[r]{\strut{}$2^{8}$}}%
      \csname LTb\endcsname
      \put(552,1641){\makebox(0,0)[r]{\strut{}$2^{10}$}}%
      \csname LTb\endcsname
      \put(552,2130){\makebox(0,0)[r]{\strut{}$2^{12}$}}%
      \csname LTb\endcsname
      \put(552,2620){\makebox(0,0)[r]{\strut{}$2^{14}$}}%
      \csname LTb\endcsname
      \put(552,3110){\makebox(0,0)[r]{\strut{}$2^{16}$}}%
      \csname LTb\endcsname
      \put(690,382){\makebox(0,0){\strut{}$1$}}%
      \csname LTb\endcsname
      \put(1833,382){\makebox(0,0){\strut{}$2$}}%
      \csname LTb\endcsname
      \put(2976,382){\makebox(0,0){\strut{}$4$}}%
      \csname LTb\endcsname
      \put(4118,382){\makebox(0,0){\strut{}$8$}}%
      \csname LTb\endcsname
      \put(5261,382){\makebox(0,0){\strut{}$16$}}%
      \put(5399,665){\makebox(0,0)[l]{\strut{}$0.199\mu s$}}%
      \put(5399,1192){\makebox(0,0)[l]{\strut{}$0.791\mu s$}}%
      \put(5399,1720){\makebox(0,0)[l]{\strut{}$3.16\mu s$}}%
      \put(5399,2247){\makebox(0,0)[l]{\strut{}$12.6\mu s$}}%
      \put(5399,2775){\makebox(0,0)[l]{\strut{}$50.5\mu s$}}%
      \put(5399,3302){\makebox(0,0)[l]{\strut{}$202\mu s$}}%
    }%
    \gplgaddtomacro\gplfronttext{%
      \csname LTb\endcsname
      \put(641,3732){\makebox(0,0)[r]{\strut{}MPI}}%
      \csname LTb\endcsname
      \put(2084,3732){\makebox(0,0)[r]{\strut{}HPX}}%
      \csname LTb\endcsname
      \put(3527,3732){\makebox(0,0)[r]{\strut{}ITO}}%
      \csname LTb\endcsname
      \put(4970,3732){\makebox(0,0)[r]{\strut{}FBC}}%
      \csname LTb\endcsname
      \put(-80,1983){\rotatebox{-270.00}{\makebox(0,0){\strut{}Kernel Iterations}}}%
      \put(6756,1983){\rotatebox{-270.00}{\makebox(0,0){\strut{}Kernel Runtime}}}%
      \put(2975,37){\makebox(0,0){\strut{}Number of Nodes}}%
    }%
    \gplbacktext
    \put(0,0){\includegraphics[width={297.60bp},height={198.40bp}]{metgWeakScaling}}%
    \gplfronttext
  \end{picture}%
\endgroup

%% file: 04fig3.tex
\begin{figure}[t]
    \centering
    \begin{subfigure}[t]{0.49\linewidth}
        \resizebox{0.99\linewidth}{!}{\input{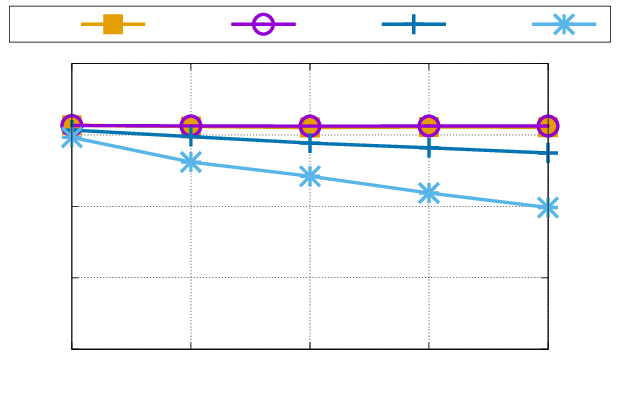}}
        \caption{Imbalance factor 0.5}
        \label{subfig:imba05}
    \end{subfigure}\hfill
    \begin{subfigure}[t]{0.49\linewidth}
        \resizebox{0.99\linewidth}{!}{\input{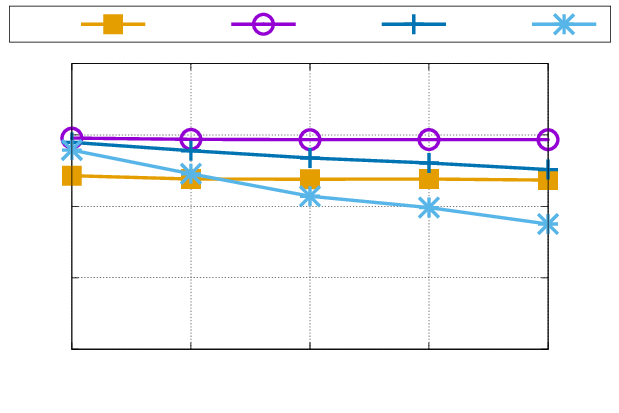}}
        \caption{Imbalance factor 1.0}
        \label{subfig:imba10}
    \end{subfigure}

    \vspace*{5pt}

    \begin{subfigure}[t]{0.49\linewidth}
        \resizebox{0.99\linewidth}{!}{\input{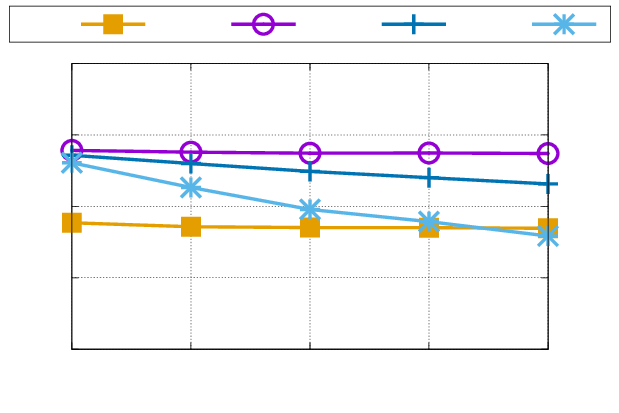}}
        \caption{Imbalance factor 1.5}
        \label{subfig:imba15}
    \end{subfigure}\hfill
    \begin{subfigure}[t]{0.49\linewidth}
        \resizebox{0.99\linewidth}{!}{\input{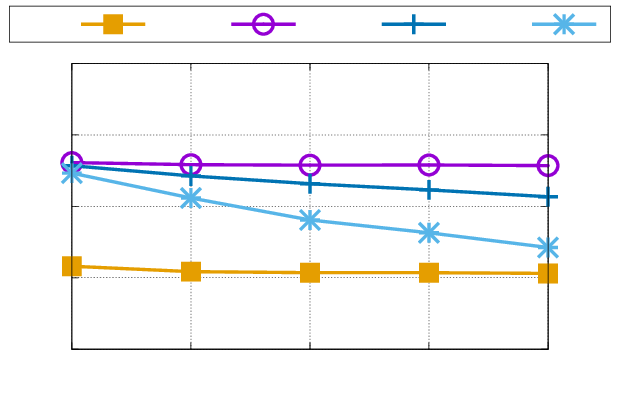}}
        \caption{Imbalance factor 2.0}
        \label{subfig:imba20}
    \end{subfigure}

    \caption{Performance with \texttt{load\_imbalance} kernels for a \texttt{stencil} graph, $2^{20}$ iterations, and 16 width per core}
    \label{fig:imba}
\end{figure}

%% file: figs/imba05.tex
\begingroup
\Large
  \makeatletter
  \providecommand\color[2][]{%
    \GenericError{(gnuplot) \space\space\space\@spaces}{%
      Package color not loaded in conjunction with
      terminal option `colourtext'%
    }{See the gnuplot documentation for explanation.%
    }{Either use 'blacktext' in gnuplot or load the package
      color.sty in LaTeX.}%
    \renewcommand\color[2][]{}%
  }%
  \providecommand\includegraphics[2][]{%
    \GenericError{(gnuplot) \space\space\space\@spaces}{%
      Package graphicx or graphics not loaded%
    }{See the gnuplot documentation for explanation.%
    }{The gnuplot epslatex terminal needs graphicx.sty or graphics.sty.}%
    \renewcommand\includegraphics[2][]{}%
  }%
  \providecommand\rotatebox[2]{#2}%
  \@ifundefined{ifGPcolor}{%
    \newif\ifGPcolor
    \GPcolortrue
  }{}%
  \@ifundefined{ifGPblacktext}{%
    \newif\ifGPblacktext
    \GPblacktextfalse
  }{}%
  \let\gplgaddtomacro\g@addto@macro
  \gdef\gplbacktext{}%
  \gdef\gplfronttext{}%
  \makeatother
  \ifGPblacktext
    \def\colorrgb#1{}%
    \def\colorgray#1{}%
  \else
    \ifGPcolor
      \def\colorrgb#1{\color[rgb]{#1}}%
      \def\colorgray#1{\color[gray]{#1}}%
      \expandafter\def\csname LTw\endcsname{\color{white}}%
      \expandafter\def\csname LTb\endcsname{\color{black}}%
      \expandafter\def\csname LTa\endcsname{\color{black}}%
      \expandafter\def\csname LT0\endcsname{\color[rgb]{1,0,0}}%
      \expandafter\def\csname LT1\endcsname{\color[rgb]{0,1,0}}%
      \expandafter\def\csname LT2\endcsname{\color[rgb]{0,0,1}}%
      \expandafter\def\csname LT3\endcsname{\color[rgb]{1,0,1}}%
      \expandafter\def\csname LT4\endcsname{\color[rgb]{0,1,1}}%
      \expandafter\def\csname LT5\endcsname{\color[rgb]{1,1,0}}%
      \expandafter\def\csname LT6\endcsname{\color[rgb]{0,0,0}}%
      \expandafter\def\csname LT7\endcsname{\color[rgb]{1,0.3,0}}%
      \expandafter\def\csname LT8\endcsname{\color[rgb]{0.5,0.5,0.5}}%
    \else
      \def\colorrgb#1{\color{black}}%
      \def\colorgray#1{\color[gray]{#1}}%
      \expandafter\def\csname LTw\endcsname{\color{white}}%
      \expandafter\def\csname LTb\endcsname{\color{black}}%
      \expandafter\def\csname LTa\endcsname{\color{black}}%
      \expandafter\def\csname LT0\endcsname{\color{black}}%
      \expandafter\def\csname LT1\endcsname{\color{black}}%
      \expandafter\def\csname LT2\endcsname{\color{black}}%
      \expandafter\def\csname LT3\endcsname{\color{black}}%
      \expandafter\def\csname LT4\endcsname{\color{black}}%
      \expandafter\def\csname LT5\endcsname{\color{black}}%
      \expandafter\def\csname LT6\endcsname{\color{black}}%
      \expandafter\def\csname LT7\endcsname{\color{black}}%
      \expandafter\def\csname LT8\endcsname{\color{black}}%
    \fi
  \fi
    \setlength{\unitlength}{0.0500bp}%
    \ifx\gptboxheight\undefined%
      \newlength{\gptboxheight}%
      \newlength{\gptboxwidth}%
      \newsavebox{\gptboxtext}%
    \fi%
    \setlength{\fboxrule}{0.5pt}%
    \setlength{\fboxsep}{1pt}%
    \definecolor{tbcol}{rgb}{1,1,1}%
\begin{picture}(5952.00,3968.00)%
    \gplgaddtomacro\gplbacktext{%
      \csname LTb\endcsname
      \put(552,612){\makebox(0,0)[r]{\strut{}80\%}}%
      \csname LTb\endcsname
      \put(552,1298){\makebox(0,0)[r]{\strut{}85\%}}%
      \csname LTb\endcsname
      \put(552,1984){\makebox(0,0)[r]{\strut{}90\%}}%
      \csname LTb\endcsname
      \put(552,2669){\makebox(0,0)[r]{\strut{}95\%}}%
      \csname LTb\endcsname
      \put(552,3355){\makebox(0,0)[r]{\strut{}100\%}}%
      \csname LTb\endcsname
      \put(690,382){\makebox(0,0){\strut{}$1$}}%
      \csname LTb\endcsname
      \put(1833,382){\makebox(0,0){\strut{}$2$}}%
      \csname LTb\endcsname
      \put(2976,382){\makebox(0,0){\strut{}$4$}}%
      \csname LTb\endcsname
      \put(4118,382){\makebox(0,0){\strut{}$8$}}%
      \csname LTb\endcsname
      \put(5261,382){\makebox(0,0){\strut{}$16$}}%
    }%
    \gplgaddtomacro\gplfronttext{%
      \csname LTb\endcsname
      \put(641,3732){\makebox(0,0)[r]{\strut{}MPI}}%
      \csname LTb\endcsname
      \put(2084,3732){\makebox(0,0)[r]{\strut{}HPX}}%
      \csname LTb\endcsname
      \put(3527,3732){\makebox(0,0)[r]{\strut{}ITO}}%
      \csname LTb\endcsname
      \put(4970,3732){\makebox(0,0)[r]{\strut{}FBC}}%
      \csname LTb\endcsname
      \put(-356,1983){\rotatebox{-270.00}{\makebox(0,0){\strut{}Efficiency}}}%
      \put(2975,37){\makebox(0,0){\strut{}Number of Nodes}}%
    }%
    \gplbacktext
    \put(0,0){\includegraphics[width={297.60bp},height={198.40bp}]{imba05}}%
    \gplfronttext
  \end{picture}%
\endgroup

%% file: figs/imba10.tex
\begingroup
\Large
  \makeatletter
  \providecommand\color[2][]{%
    \GenericError{(gnuplot) \space\space\space\@spaces}{%
      Package color not loaded in conjunction with
      terminal option `colourtext'%
    }{See the gnuplot documentation for explanation.%
    }{Either use 'blacktext' in gnuplot or load the package
      color.sty in LaTeX.}%
    \renewcommand\color[2][]{}%
  }%
  \providecommand\includegraphics[2][]{%
    \GenericError{(gnuplot) \space\space\space\@spaces}{%
      Package graphicx or graphics not loaded%
    }{See the gnuplot documentation for explanation.%
    }{The gnuplot epslatex terminal needs graphicx.sty or graphics.sty.}%
    \renewcommand\includegraphics[2][]{}%
  }%
  \providecommand\rotatebox[2]{#2}%
  \@ifundefined{ifGPcolor}{%
    \newif\ifGPcolor
    \GPcolortrue
  }{}%
  \@ifundefined{ifGPblacktext}{%
    \newif\ifGPblacktext
    \GPblacktextfalse
  }{}%
  \let\gplgaddtomacro\g@addto@macro
  \gdef\gplbacktext{}%
  \gdef\gplfronttext{}%
  \makeatother
  \ifGPblacktext
    \def\colorrgb#1{}%
    \def\colorgray#1{}%
  \else
    \ifGPcolor
      \def\colorrgb#1{\color[rgb]{#1}}%
      \def\colorgray#1{\color[gray]{#1}}%
      \expandafter\def\csname LTw\endcsname{\color{white}}%
      \expandafter\def\csname LTb\endcsname{\color{black}}%
      \expandafter\def\csname LTa\endcsname{\color{black}}%
      \expandafter\def\csname LT0\endcsname{\color[rgb]{1,0,0}}%
      \expandafter\def\csname LT1\endcsname{\color[rgb]{0,1,0}}%
      \expandafter\def\csname LT2\endcsname{\color[rgb]{0,0,1}}%
      \expandafter\def\csname LT3\endcsname{\color[rgb]{1,0,1}}%
      \expandafter\def\csname LT4\endcsname{\color[rgb]{0,1,1}}%
      \expandafter\def\csname LT5\endcsname{\color[rgb]{1,1,0}}%
      \expandafter\def\csname LT6\endcsname{\color[rgb]{0,0,0}}%
      \expandafter\def\csname LT7\endcsname{\color[rgb]{1,0.3,0}}%
      \expandafter\def\csname LT8\endcsname{\color[rgb]{0.5,0.5,0.5}}%
    \else
      \def\colorrgb#1{\color{black}}%
      \def\colorgray#1{\color[gray]{#1}}%
      \expandafter\def\csname LTw\endcsname{\color{white}}%
      \expandafter\def\csname LTb\endcsname{\color{black}}%
      \expandafter\def\csname LTa\endcsname{\color{black}}%
      \expandafter\def\csname LT0\endcsname{\color{black}}%
      \expandafter\def\csname LT1\endcsname{\color{black}}%
      \expandafter\def\csname LT2\endcsname{\color{black}}%
      \expandafter\def\csname LT3\endcsname{\color{black}}%
      \expandafter\def\csname LT4\endcsname{\color{black}}%
      \expandafter\def\csname LT5\endcsname{\color{black}}%
      \expandafter\def\csname LT6\endcsname{\color{black}}%
      \expandafter\def\csname LT7\endcsname{\color{black}}%
      \expandafter\def\csname LT8\endcsname{\color{black}}%
    \fi
  \fi
    \setlength{\unitlength}{0.0500bp}%
    \ifx\gptboxheight\undefined%
      \newlength{\gptboxheight}%
      \newlength{\gptboxwidth}%
      \newsavebox{\gptboxtext}%
    \fi%
    \setlength{\fboxrule}{0.5pt}%
    \setlength{\fboxsep}{1pt}%
    \definecolor{tbcol}{rgb}{1,1,1}%
\begin{picture}(5952.00,3968.00)%
    \gplgaddtomacro\gplbacktext{%
      \csname LTb\endcsname
      \put(552,612){\makebox(0,0)[r]{\strut{}80\%}}%
      \csname LTb\endcsname
      \put(552,1298){\makebox(0,0)[r]{\strut{}85\%}}%
      \csname LTb\endcsname
      \put(552,1984){\makebox(0,0)[r]{\strut{}90\%}}%
      \csname LTb\endcsname
      \put(552,2669){\makebox(0,0)[r]{\strut{}95\%}}%
      \csname LTb\endcsname
      \put(552,3355){\makebox(0,0)[r]{\strut{}100\%}}%
      \csname LTb\endcsname
      \put(690,382){\makebox(0,0){\strut{}$1$}}%
      \csname LTb\endcsname
      \put(1833,382){\makebox(0,0){\strut{}$2$}}%
      \csname LTb\endcsname
      \put(2976,382){\makebox(0,0){\strut{}$4$}}%
      \csname LTb\endcsname
      \put(4118,382){\makebox(0,0){\strut{}$8$}}%
      \csname LTb\endcsname
      \put(5261,382){\makebox(0,0){\strut{}$16$}}%
    }%
    \gplgaddtomacro\gplfronttext{%
      \csname LTb\endcsname
      \put(641,3732){\makebox(0,0)[r]{\strut{}MPI}}%
      \csname LTb\endcsname
      \put(2084,3732){\makebox(0,0)[r]{\strut{}HPX}}%
      \csname LTb\endcsname
      \put(3527,3732){\makebox(0,0)[r]{\strut{}ITO}}%
      \csname LTb\endcsname
      \put(4970,3732){\makebox(0,0)[r]{\strut{}FBC}}%
      \csname LTb\endcsname
      \put(-356,1983){\rotatebox{-270.00}{\makebox(0,0){\strut{}Efficiency}}}%
      \put(2975,37){\makebox(0,0){\strut{}Number of Nodes}}%
    }%
    \gplbacktext
    \put(0,0){\includegraphics[width={297.60bp},height={198.40bp}]{imba10}}%
    \gplfronttext
  \end{picture}%
\endgroup

%% file: figs/imba15.tex
\begingroup
\Large
  \makeatletter
  \providecommand\color[2][]{%
    \GenericError{(gnuplot) \space\space\space\@spaces}{%
      Package color not loaded in conjunction with
      terminal option `colourtext'%
    }{See the gnuplot documentation for explanation.%
    }{Either use 'blacktext' in gnuplot or load the package
      color.sty in LaTeX.}%
    \renewcommand\color[2][]{}%
  }%
  \providecommand\includegraphics[2][]{%
    \GenericError{(gnuplot) \space\space\space\@spaces}{%
      Package graphicx or graphics not loaded%
    }{See the gnuplot documentation for explanation.%
    }{The gnuplot epslatex terminal needs graphicx.sty or graphics.sty.}%
    \renewcommand\includegraphics[2][]{}%
  }%
  \providecommand\rotatebox[2]{#2}%
  \@ifundefined{ifGPcolor}{%
    \newif\ifGPcolor
    \GPcolortrue
  }{}%
  \@ifundefined{ifGPblacktext}{%
    \newif\ifGPblacktext
    \GPblacktextfalse
  }{}%
  \let\gplgaddtomacro\g@addto@macro
  \gdef\gplbacktext{}%
  \gdef\gplfronttext{}%
  \makeatother
  \ifGPblacktext
    \def\colorrgb#1{}%
    \def\colorgray#1{}%
  \else
    \ifGPcolor
      \def\colorrgb#1{\color[rgb]{#1}}%
      \def\colorgray#1{\color[gray]{#1}}%
      \expandafter\def\csname LTw\endcsname{\color{white}}%
      \expandafter\def\csname LTb\endcsname{\color{black}}%
      \expandafter\def\csname LTa\endcsname{\color{black}}%
      \expandafter\def\csname LT0\endcsname{\color[rgb]{1,0,0}}%
      \expandafter\def\csname LT1\endcsname{\color[rgb]{0,1,0}}%
      \expandafter\def\csname LT2\endcsname{\color[rgb]{0,0,1}}%
      \expandafter\def\csname LT3\endcsname{\color[rgb]{1,0,1}}%
      \expandafter\def\csname LT4\endcsname{\color[rgb]{0,1,1}}%
      \expandafter\def\csname LT5\endcsname{\color[rgb]{1,1,0}}%
      \expandafter\def\csname LT6\endcsname{\color[rgb]{0,0,0}}%
      \expandafter\def\csname LT7\endcsname{\color[rgb]{1,0.3,0}}%
      \expandafter\def\csname LT8\endcsname{\color[rgb]{0.5,0.5,0.5}}%
    \else
      \def\colorrgb#1{\color{black}}%
      \def\colorgray#1{\color[gray]{#1}}%
      \expandafter\def\csname LTw\endcsname{\color{white}}%
      \expandafter\def\csname LTb\endcsname{\color{black}}%
      \expandafter\def\csname LTa\endcsname{\color{black}}%
      \expandafter\def\csname LT0\endcsname{\color{black}}%
      \expandafter\def\csname LT1\endcsname{\color{black}}%
      \expandafter\def\csname LT2\endcsname{\color{black}}%
      \expandafter\def\csname LT3\endcsname{\color{black}}%
      \expandafter\def\csname LT4\endcsname{\color{black}}%
      \expandafter\def\csname LT5\endcsname{\color{black}}%
      \expandafter\def\csname LT6\endcsname{\color{black}}%
      \expandafter\def\csname LT7\endcsname{\color{black}}%
      \expandafter\def\csname LT8\endcsname{\color{black}}%
    \fi
  \fi
    \setlength{\unitlength}{0.0500bp}%
    \ifx\gptboxheight\undefined%
      \newlength{\gptboxheight}%
      \newlength{\gptboxwidth}%
      \newsavebox{\gptboxtext}%
    \fi%
    \setlength{\fboxrule}{0.5pt}%
    \setlength{\fboxsep}{1pt}%
    \definecolor{tbcol}{rgb}{1,1,1}%
\begin{picture}(5952.00,3968.00)%
    \gplgaddtomacro\gplbacktext{%
      \csname LTb\endcsname
      \put(552,612){\makebox(0,0)[r]{\strut{}80\%}}%
      \csname LTb\endcsname
      \put(552,1298){\makebox(0,0)[r]{\strut{}85\%}}%
      \csname LTb\endcsname
      \put(552,1984){\makebox(0,0)[r]{\strut{}90\%}}%
      \csname LTb\endcsname
      \put(552,2669){\makebox(0,0)[r]{\strut{}95\%}}%
      \csname LTb\endcsname
      \put(552,3355){\makebox(0,0)[r]{\strut{}100\%}}%
      \csname LTb\endcsname
      \put(690,382){\makebox(0,0){\strut{}$1$}}%
      \csname LTb\endcsname
      \put(1833,382){\makebox(0,0){\strut{}$2$}}%
      \csname LTb\endcsname
      \put(2976,382){\makebox(0,0){\strut{}$4$}}%
      \csname LTb\endcsname
      \put(4118,382){\makebox(0,0){\strut{}$8$}}%
      \csname LTb\endcsname
      \put(5261,382){\makebox(0,0){\strut{}$16$}}%
    }%
    \gplgaddtomacro\gplfronttext{%
      \csname LTb\endcsname
      \put(641,3732){\makebox(0,0)[r]{\strut{}MPI}}%
      \csname LTb\endcsname
      \put(2084,3732){\makebox(0,0)[r]{\strut{}HPX}}%
      \csname LTb\endcsname
      \put(3527,3732){\makebox(0,0)[r]{\strut{}ITO}}%
      \csname LTb\endcsname
      \put(4970,3732){\makebox(0,0)[r]{\strut{}FBC}}%
      \csname LTb\endcsname
      \put(-356,1983){\rotatebox{-270.00}{\makebox(0,0){\strut{}Efficiency}}}%
      \put(2975,37){\makebox(0,0){\strut{}Number of Nodes}}%
    }%
    \gplbacktext
    \put(0,0){\includegraphics[width={297.60bp},height={198.40bp}]{imba15}}%
    \gplfronttext
  \end{picture}%
\endgroup

%% file: figs/imba20.tex
\begingroup
\Large
  \makeatletter
  \providecommand\color[2][]{%
    \GenericError{(gnuplot) \space\space\space\@spaces}{%
      Package color not loaded in conjunction with
      terminal option `colourtext'%
    }{See the gnuplot documentation for explanation.%
    }{Either use 'blacktext' in gnuplot or load the package
      color.sty in LaTeX.}%
    \renewcommand\color[2][]{}%
  }%
  \providecommand\includegraphics[2][]{%
    \GenericError{(gnuplot) \space\space\space\@spaces}{%
      Package graphicx or graphics not loaded%
    }{See the gnuplot documentation for explanation.%
    }{The gnuplot epslatex terminal needs graphicx.sty or graphics.sty.}%
    \renewcommand\includegraphics[2][]{}%
  }%
  \providecommand\rotatebox[2]{#2}%
  \@ifundefined{ifGPcolor}{%
    \newif\ifGPcolor
    \GPcolortrue
  }{}%
  \@ifundefined{ifGPblacktext}{%
    \newif\ifGPblacktext
    \GPblacktextfalse
  }{}%
  \let\gplgaddtomacro\g@addto@macro
  \gdef\gplbacktext{}%
  \gdef\gplfronttext{}%
  \makeatother
  \ifGPblacktext
    \def\colorrgb#1{}%
    \def\colorgray#1{}%
  \else
    \ifGPcolor
      \def\colorrgb#1{\color[rgb]{#1}}%
      \def\colorgray#1{\color[gray]{#1}}%
      \expandafter\def\csname LTw\endcsname{\color{white}}%
      \expandafter\def\csname LTb\endcsname{\color{black}}%
      \expandafter\def\csname LTa\endcsname{\color{black}}%
      \expandafter\def\csname LT0\endcsname{\color[rgb]{1,0,0}}%
      \expandafter\def\csname LT1\endcsname{\color[rgb]{0,1,0}}%
      \expandafter\def\csname LT2\endcsname{\color[rgb]{0,0,1}}%
      \expandafter\def\csname LT3\endcsname{\color[rgb]{1,0,1}}%
      \expandafter\def\csname LT4\endcsname{\color[rgb]{0,1,1}}%
      \expandafter\def\csname LT5\endcsname{\color[rgb]{1,1,0}}%
      \expandafter\def\csname LT6\endcsname{\color[rgb]{0,0,0}}%
      \expandafter\def\csname LT7\endcsname{\color[rgb]{1,0.3,0}}%
      \expandafter\def\csname LT8\endcsname{\color[rgb]{0.5,0.5,0.5}}%
    \else
      \def\colorrgb#1{\color{black}}%
      \def\colorgray#1{\color[gray]{#1}}%
      \expandafter\def\csname LTw\endcsname{\color{white}}%
      \expandafter\def\csname LTb\endcsname{\color{black}}%
      \expandafter\def\csname LTa\endcsname{\color{black}}%
      \expandafter\def\csname LT0\endcsname{\color{black}}%
      \expandafter\def\csname LT1\endcsname{\color{black}}%
      \expandafter\def\csname LT2\endcsname{\color{black}}%
      \expandafter\def\csname LT3\endcsname{\color{black}}%
      \expandafter\def\csname LT4\endcsname{\color{black}}%
      \expandafter\def\csname LT5\endcsname{\color{black}}%
      \expandafter\def\csname LT6\endcsname{\color{black}}%
      \expandafter\def\csname LT7\endcsname{\color{black}}%
      \expandafter\def\csname LT8\endcsname{\color{black}}%
    \fi
  \fi
    \setlength{\unitlength}{0.0500bp}%
    \ifx\gptboxheight\undefined%
      \newlength{\gptboxheight}%
      \newlength{\gptboxwidth}%
      \newsavebox{\gptboxtext}%
    \fi%
    \setlength{\fboxrule}{0.5pt}%
    \setlength{\fboxsep}{1pt}%
    \definecolor{tbcol}{rgb}{1,1,1}%
\begin{picture}(5952.00,3968.00)%
    \gplgaddtomacro\gplbacktext{%
      \csname LTb\endcsname
      \put(552,612){\makebox(0,0)[r]{\strut{}80\%}}%
      \csname LTb\endcsname
      \put(552,1298){\makebox(0,0)[r]{\strut{}85\%}}%
      \csname LTb\endcsname
      \put(552,1984){\makebox(0,0)[r]{\strut{}90\%}}%
      \csname LTb\endcsname
      \put(552,2669){\makebox(0,0)[r]{\strut{}95\%}}%
      \csname LTb\endcsname
      \put(552,3355){\makebox(0,0)[r]{\strut{}100\%}}%
      \csname LTb\endcsname
      \put(690,382){\makebox(0,0){\strut{}$1$}}%
      \csname LTb\endcsname
      \put(1833,382){\makebox(0,0){\strut{}$2$}}%
      \csname LTb\endcsname
      \put(2976,382){\makebox(0,0){\strut{}$4$}}%
      \csname LTb\endcsname
      \put(4118,382){\makebox(0,0){\strut{}$8$}}%
      \csname LTb\endcsname
      \put(5261,382){\makebox(0,0){\strut{}$16$}}%
    }%
    \gplgaddtomacro\gplfronttext{%
      \csname LTb\endcsname
      \put(641,3732){\makebox(0,0)[r]{\strut{}MPI}}%
      \csname LTb\endcsname
      \put(2084,3732){\makebox(0,0)[r]{\strut{}HPX}}%
      \csname LTb\endcsname
      \put(3527,3732){\makebox(0,0)[r]{\strut{}ITO}}%
      \csname LTb\endcsname
      \put(4970,3732){\makebox(0,0)[r]{\strut{}FBC}}%
      \csname LTb\endcsname
      \put(-356,1983){\rotatebox{-270.00}{\makebox(0,0){\strut{}Efficiency}}}%
      \put(2975,37){\makebox(0,0){\strut{}Number of Nodes}}%
    }%
    \gplbacktext
    \put(0,0){\includegraphics[width={297.60bp},height={198.40bp}]{imba20}}%
    \gplfronttext
  \end{picture}%
\endgroup

%% file: 04fig4.tex
\begin{figure}[t]
    \centering
    \begin{subfigure}[t]{0.49\linewidth}
        \resizebox{0.99\linewidth}{!}{\input{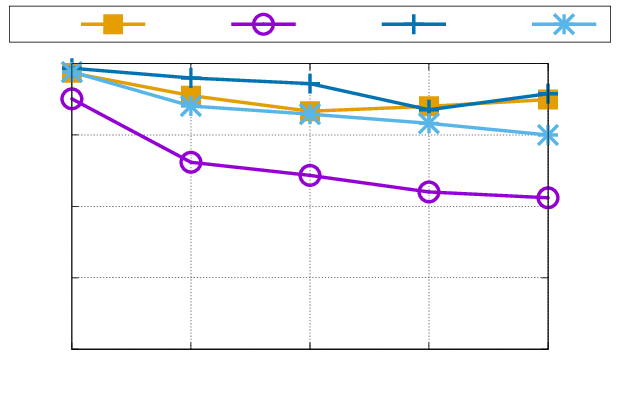}}
        \caption{40 dependencies per task}
        \label{subfig:spread40}
    \end{subfigure}\hfill
    \begin{subfigure}[t]{0.49\linewidth}
        \resizebox{0.99\linewidth}{!}{\input{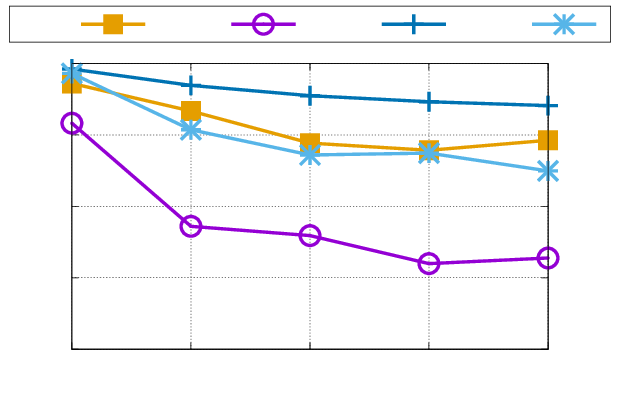}}
        \caption{80 dependencies per task}
        \label{subfig:spread80}
    \end{subfigure}

    \caption{Communication-focused experiment with a \texttt{spread} graph, \texttt{compute\_bound} kernel, $2^{20}$ iterations, 1\,KiB output size, and 16 width per core}
    \label{fig:spread}
\end{figure}

%% file: figs/spread40.tex
\begingroup
\Large
  \makeatletter
  \providecommand\color[2][]{%
    \GenericError{(gnuplot) \space\space\space\@spaces}{%
      Package color not loaded in conjunction with
      terminal option `colourtext'%
    }{See the gnuplot documentation for explanation.%
    }{Either use 'blacktext' in gnuplot or load the package
      color.sty in LaTeX.}%
    \renewcommand\color[2][]{}%
  }%
  \providecommand\includegraphics[2][]{%
    \GenericError{(gnuplot) \space\space\space\@spaces}{%
      Package graphicx or graphics not loaded%
    }{See the gnuplot documentation for explanation.%
    }{The gnuplot epslatex terminal needs graphicx.sty or graphics.sty.}%
    \renewcommand\includegraphics[2][]{}%
  }%
  \providecommand\rotatebox[2]{#2}%
  \@ifundefined{ifGPcolor}{%
    \newif\ifGPcolor
    \GPcolortrue
  }{}%
  \@ifundefined{ifGPblacktext}{%
    \newif\ifGPblacktext
    \GPblacktextfalse
  }{}%
  \let\gplgaddtomacro\g@addto@macro
  \gdef\gplbacktext{}%
  \gdef\gplfronttext{}%
  \makeatother
  \ifGPblacktext
    \def\colorrgb#1{}%
    \def\colorgray#1{}%
  \else
    \ifGPcolor
      \def\colorrgb#1{\color[rgb]{#1}}%
      \def\colorgray#1{\color[gray]{#1}}%
      \expandafter\def\csname LTw\endcsname{\color{white}}%
      \expandafter\def\csname LTb\endcsname{\color{black}}%
      \expandafter\def\csname LTa\endcsname{\color{black}}%
      \expandafter\def\csname LT0\endcsname{\color[rgb]{1,0,0}}%
      \expandafter\def\csname LT1\endcsname{\color[rgb]{0,1,0}}%
      \expandafter\def\csname LT2\endcsname{\color[rgb]{0,0,1}}%
      \expandafter\def\csname LT3\endcsname{\color[rgb]{1,0,1}}%
      \expandafter\def\csname LT4\endcsname{\color[rgb]{0,1,1}}%
      \expandafter\def\csname LT5\endcsname{\color[rgb]{1,1,0}}%
      \expandafter\def\csname LT6\endcsname{\color[rgb]{0,0,0}}%
      \expandafter\def\csname LT7\endcsname{\color[rgb]{1,0.3,0}}%
      \expandafter\def\csname LT8\endcsname{\color[rgb]{0.5,0.5,0.5}}%
    \else
      \def\colorrgb#1{\color{black}}%
      \def\colorgray#1{\color[gray]{#1}}%
      \expandafter\def\csname LTw\endcsname{\color{white}}%
      \expandafter\def\csname LTb\endcsname{\color{black}}%
      \expandafter\def\csname LTa\endcsname{\color{black}}%
      \expandafter\def\csname LT0\endcsname{\color{black}}%
      \expandafter\def\csname LT1\endcsname{\color{black}}%
      \expandafter\def\csname LT2\endcsname{\color{black}}%
      \expandafter\def\csname LT3\endcsname{\color{black}}%
      \expandafter\def\csname LT4\endcsname{\color{black}}%
      \expandafter\def\csname LT5\endcsname{\color{black}}%
      \expandafter\def\csname LT6\endcsname{\color{black}}%
      \expandafter\def\csname LT7\endcsname{\color{black}}%
      \expandafter\def\csname LT8\endcsname{\color{black}}%
    \fi
  \fi
    \setlength{\unitlength}{0.0500bp}%
    \ifx\gptboxheight\undefined%
      \newlength{\gptboxheight}%
      \newlength{\gptboxwidth}%
      \newsavebox{\gptboxtext}%
    \fi%
    \setlength{\fboxrule}{0.5pt}%
    \setlength{\fboxsep}{1pt}%
    \definecolor{tbcol}{rgb}{1,1,1}%
\begin{picture}(5952.00,3968.00)%
    \gplgaddtomacro\gplbacktext{%
      \csname LTb\endcsname
      \put(552,612){\makebox(0,0)[r]{\strut{}0\%}}%
      \csname LTb\endcsname
      \put(552,1298){\makebox(0,0)[r]{\strut{}25\%}}%
      \csname LTb\endcsname
      \put(552,1984){\makebox(0,0)[r]{\strut{}50\%}}%
      \csname LTb\endcsname
      \put(552,2669){\makebox(0,0)[r]{\strut{}75\%}}%
      \csname LTb\endcsname
      \put(552,3355){\makebox(0,0)[r]{\strut{}100\%}}%
      \csname LTb\endcsname
      \put(690,382){\makebox(0,0){\strut{}$1$}}%
      \csname LTb\endcsname
      \put(1833,382){\makebox(0,0){\strut{}$2$}}%
      \csname LTb\endcsname
      \put(2976,382){\makebox(0,0){\strut{}$4$}}%
      \csname LTb\endcsname
      \put(4118,382){\makebox(0,0){\strut{}$8$}}%
      \csname LTb\endcsname
      \put(5261,382){\makebox(0,0){\strut{}$16$}}%
    }%
    \gplgaddtomacro\gplfronttext{%
      \csname LTb\endcsname
      \put(641,3732){\makebox(0,0)[r]{\strut{}MPI}}%
      \csname LTb\endcsname
      \put(2084,3732){\makebox(0,0)[r]{\strut{}HPX}}%
      \csname LTb\endcsname
      \put(3527,3732){\makebox(0,0)[r]{\strut{}ITO}}%
      \csname LTb\endcsname
      \put(4970,3732){\makebox(0,0)[r]{\strut{}FBC}}%
      \csname LTb\endcsname
      \put(-356,1983){\rotatebox{-270.00}{\makebox(0,0){\strut{}Efficiency}}}%
      \put(2975,37){\makebox(0,0){\strut{}Number of Nodes}}%
    }%
    \gplbacktext
    \put(0,0){\includegraphics[width={297.60bp},height={198.40bp}]{spread40}}%
    \gplfronttext
  \end{picture}%
\endgroup

%% file: figs/spread80.tex
\begingroup
\Large
  \makeatletter
  \providecommand\color[2][]{%
    \GenericError{(gnuplot) \space\space\space\@spaces}{%
      Package color not loaded in conjunction with
      terminal option `colourtext'%
    }{See the gnuplot documentation for explanation.%
    }{Either use 'blacktext' in gnuplot or load the package
      color.sty in LaTeX.}%
    \renewcommand\color[2][]{}%
  }%
  \providecommand\includegraphics[2][]{%
    \GenericError{(gnuplot) \space\space\space\@spaces}{%
      Package graphicx or graphics not loaded%
    }{See the gnuplot documentation for explanation.%
    }{The gnuplot epslatex terminal needs graphicx.sty or graphics.sty.}%
    \renewcommand\includegraphics[2][]{}%
  }%
  \providecommand\rotatebox[2]{#2}%
  \@ifundefined{ifGPcolor}{%
    \newif\ifGPcolor
    \GPcolortrue
  }{}%
  \@ifundefined{ifGPblacktext}{%
    \newif\ifGPblacktext
    \GPblacktextfalse
  }{}%
  \let\gplgaddtomacro\g@addto@macro
  \gdef\gplbacktext{}%
  \gdef\gplfronttext{}%
  \makeatother
  \ifGPblacktext
    \def\colorrgb#1{}%
    \def\colorgray#1{}%
  \else
    \ifGPcolor
      \def\colorrgb#1{\color[rgb]{#1}}%
      \def\colorgray#1{\color[gray]{#1}}%
      \expandafter\def\csname LTw\endcsname{\color{white}}%
      \expandafter\def\csname LTb\endcsname{\color{black}}%
      \expandafter\def\csname LTa\endcsname{\color{black}}%
      \expandafter\def\csname LT0\endcsname{\color[rgb]{1,0,0}}%
      \expandafter\def\csname LT1\endcsname{\color[rgb]{0,1,0}}%
      \expandafter\def\csname LT2\endcsname{\color[rgb]{0,0,1}}%
      \expandafter\def\csname LT3\endcsname{\color[rgb]{1,0,1}}%
      \expandafter\def\csname LT4\endcsname{\color[rgb]{0,1,1}}%
      \expandafter\def\csname LT5\endcsname{\color[rgb]{1,1,0}}%
      \expandafter\def\csname LT6\endcsname{\color[rgb]{0,0,0}}%
      \expandafter\def\csname LT7\endcsname{\color[rgb]{1,0.3,0}}%
      \expandafter\def\csname LT8\endcsname{\color[rgb]{0.5,0.5,0.5}}%
    \else
      \def\colorrgb#1{\color{black}}%
      \def\colorgray#1{\color[gray]{#1}}%
      \expandafter\def\csname LTw\endcsname{\color{white}}%
      \expandafter\def\csname LTb\endcsname{\color{black}}%
      \expandafter\def\csname LTa\endcsname{\color{black}}%
      \expandafter\def\csname LT0\endcsname{\color{black}}%
      \expandafter\def\csname LT1\endcsname{\color{black}}%
      \expandafter\def\csname LT2\endcsname{\color{black}}%
      \expandafter\def\csname LT3\endcsname{\color{black}}%
      \expandafter\def\csname LT4\endcsname{\color{black}}%
      \expandafter\def\csname LT5\endcsname{\color{black}}%
      \expandafter\def\csname LT6\endcsname{\color{black}}%
      \expandafter\def\csname LT7\endcsname{\color{black}}%
      \expandafter\def\csname LT8\endcsname{\color{black}}%
    \fi
  \fi
    \setlength{\unitlength}{0.0500bp}%
    \ifx\gptboxheight\undefined%
      \newlength{\gptboxheight}%
      \newlength{\gptboxwidth}%
      \newsavebox{\gptboxtext}%
    \fi%
    \setlength{\fboxrule}{0.5pt}%
    \setlength{\fboxsep}{1pt}%
    \definecolor{tbcol}{rgb}{1,1,1}%
\begin{picture}(5952.00,3968.00)%
    \gplgaddtomacro\gplbacktext{%
      \csname LTb\endcsname
      \put(552,612){\makebox(0,0)[r]{\strut{}0\%}}%
      \csname LTb\endcsname
      \put(552,1298){\makebox(0,0)[r]{\strut{}25\%}}%
      \csname LTb\endcsname
      \put(552,1984){\makebox(0,0)[r]{\strut{}50\%}}%
      \csname LTb\endcsname
      \put(552,2669){\makebox(0,0)[r]{\strut{}75\%}}%
      \csname LTb\endcsname
      \put(552,3355){\makebox(0,0)[r]{\strut{}100\%}}%
      \csname LTb\endcsname
      \put(690,382){\makebox(0,0){\strut{}$1$}}%
      \csname LTb\endcsname
      \put(1833,382){\makebox(0,0){\strut{}$2$}}%
      \csname LTb\endcsname
      \put(2976,382){\makebox(0,0){\strut{}$4$}}%
      \csname LTb\endcsname
      \put(4118,382){\makebox(0,0){\strut{}$8$}}%
      \csname LTb\endcsname
      \put(5261,382){\makebox(0,0){\strut{}$16$}}%
    }%
    \gplgaddtomacro\gplfronttext{%
      \csname LTb\endcsname
      \put(641,3732){\makebox(0,0)[r]{\strut{}MPI}}%
      \csname LTb\endcsname
      \put(2084,3732){\makebox(0,0)[r]{\strut{}HPX}}%
      \csname LTb\endcsname
      \put(3527,3732){\makebox(0,0)[r]{\strut{}ITO}}%
      \csname LTb\endcsname
      \put(4970,3732){\makebox(0,0)[r]{\strut{}FBC}}%
      \csname LTb\endcsname
      \put(-356,1983){\rotatebox{-270.00}{\makebox(0,0){\strut{}Efficiency}}}%
      \put(2975,37){\makebox(0,0){\strut{}Number of Nodes}}%
    }%
    \gplbacktext
    \put(0,0){\includegraphics[width={297.60bp},height={198.40bp}]{spread80}}%
    \gplfronttext
  \end{picture}%
\endgroup

%% file: 04fig5.tex
\begin{figure}[t]
    \centering
    \begin{subfigure}[t]{0.49\linewidth}
        \resizebox{0.99\linewidth}{!}{\input{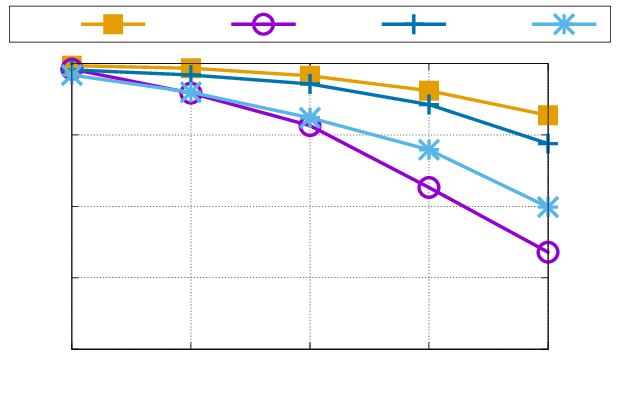}}
        \caption{1 width per core}
        \label{subfig:ata1}
    \end{subfigure}\hfill
    \begin{subfigure}[t]{0.49\linewidth}
        \resizebox{0.99\linewidth}{!}{\input{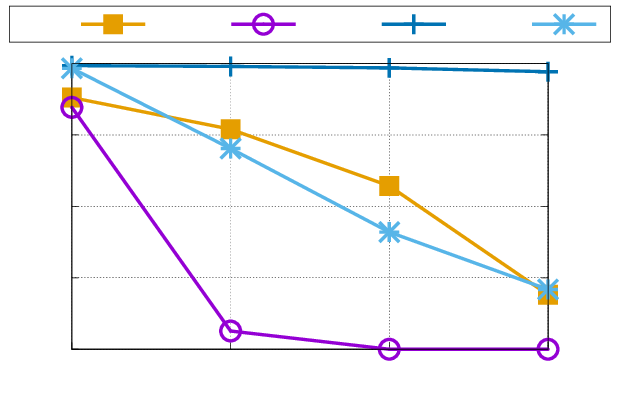}}
        \caption{16 width per core}
        \label{subfig:ata16}
    \end{subfigure}

    \caption{Communication-focused experiment with an \texttt{all\_to\_all} graph, \texttt{compute\_bound} kernel, $2^{20}$ iterations, and 16\,B output size}
    \label{fig:ata}
\end{figure}

%% file: figs/ata1.tex
\begingroup
\Large
  \makeatletter
  \providecommand\color[2][]{%
    \GenericError{(gnuplot) \space\space\space\@spaces}{%
      Package color not loaded in conjunction with
      terminal option `colourtext'%
    }{See the gnuplot documentation for explanation.%
    }{Either use 'blacktext' in gnuplot or load the package
      color.sty in LaTeX.}%
    \renewcommand\color[2][]{}%
  }%
  \providecommand\includegraphics[2][]{%
    \GenericError{(gnuplot) \space\space\space\@spaces}{%
      Package graphicx or graphics not loaded%
    }{See the gnuplot documentation for explanation.%
    }{The gnuplot epslatex terminal needs graphicx.sty or graphics.sty.}%
    \renewcommand\includegraphics[2][]{}%
  }%
  \providecommand\rotatebox[2]{#2}%
  \@ifundefined{ifGPcolor}{%
    \newif\ifGPcolor
    \GPcolortrue
  }{}%
  \@ifundefined{ifGPblacktext}{%
    \newif\ifGPblacktext
    \GPblacktextfalse
  }{}%
  \let\gplgaddtomacro\g@addto@macro
  \gdef\gplbacktext{}%
  \gdef\gplfronttext{}%
  \makeatother
  \ifGPblacktext
    \def\colorrgb#1{}%
    \def\colorgray#1{}%
  \else
    \ifGPcolor
      \def\colorrgb#1{\color[rgb]{#1}}%
      \def\colorgray#1{\color[gray]{#1}}%
      \expandafter\def\csname LTw\endcsname{\color{white}}%
      \expandafter\def\csname LTb\endcsname{\color{black}}%
      \expandafter\def\csname LTa\endcsname{\color{black}}%
      \expandafter\def\csname LT0\endcsname{\color[rgb]{1,0,0}}%
      \expandafter\def\csname LT1\endcsname{\color[rgb]{0,1,0}}%
      \expandafter\def\csname LT2\endcsname{\color[rgb]{0,0,1}}%
      \expandafter\def\csname LT3\endcsname{\color[rgb]{1,0,1}}%
      \expandafter\def\csname LT4\endcsname{\color[rgb]{0,1,1}}%
      \expandafter\def\csname LT5\endcsname{\color[rgb]{1,1,0}}%
      \expandafter\def\csname LT6\endcsname{\color[rgb]{0,0,0}}%
      \expandafter\def\csname LT7\endcsname{\color[rgb]{1,0.3,0}}%
      \expandafter\def\csname LT8\endcsname{\color[rgb]{0.5,0.5,0.5}}%
    \else
      \def\colorrgb#1{\color{black}}%
      \def\colorgray#1{\color[gray]{#1}}%
      \expandafter\def\csname LTw\endcsname{\color{white}}%
      \expandafter\def\csname LTb\endcsname{\color{black}}%
      \expandafter\def\csname LTa\endcsname{\color{black}}%
      \expandafter\def\csname LT0\endcsname{\color{black}}%
      \expandafter\def\csname LT1\endcsname{\color{black}}%
      \expandafter\def\csname LT2\endcsname{\color{black}}%
      \expandafter\def\csname LT3\endcsname{\color{black}}%
      \expandafter\def\csname LT4\endcsname{\color{black}}%
      \expandafter\def\csname LT5\endcsname{\color{black}}%
      \expandafter\def\csname LT6\endcsname{\color{black}}%
      \expandafter\def\csname LT7\endcsname{\color{black}}%
      \expandafter\def\csname LT8\endcsname{\color{black}}%
    \fi
  \fi
    \setlength{\unitlength}{0.0500bp}%
    \ifx\gptboxheight\undefined%
      \newlength{\gptboxheight}%
      \newlength{\gptboxwidth}%
      \newsavebox{\gptboxtext}%
    \fi%
    \setlength{\fboxrule}{0.5pt}%
    \setlength{\fboxsep}{1pt}%
    \definecolor{tbcol}{rgb}{1,1,1}%
\begin{picture}(5952.00,3968.00)%
    \gplgaddtomacro\gplbacktext{%
      \csname LTb\endcsname
      \put(552,612){\makebox(0,0)[r]{\strut{}0\%}}%
      \csname LTb\endcsname
      \put(552,1298){\makebox(0,0)[r]{\strut{}25\%}}%
      \csname LTb\endcsname
      \put(552,1984){\makebox(0,0)[r]{\strut{}50\%}}%
      \csname LTb\endcsname
      \put(552,2669){\makebox(0,0)[r]{\strut{}75\%}}%
      \csname LTb\endcsname
      \put(552,3355){\makebox(0,0)[r]{\strut{}100\%}}%
      \csname LTb\endcsname
      \put(690,382){\makebox(0,0){\strut{}$1$}}%
      \csname LTb\endcsname
      \put(1833,382){\makebox(0,0){\strut{}$2$}}%
      \csname LTb\endcsname
      \put(2976,382){\makebox(0,0){\strut{}$4$}}%
      \csname LTb\endcsname
      \put(4118,382){\makebox(0,0){\strut{}$8$}}%
      \csname LTb\endcsname
      \put(5261,382){\makebox(0,0){\strut{}$16$}}%
    }%
    \gplgaddtomacro\gplfronttext{%
      \csname LTb\endcsname
      \put(641,3732){\makebox(0,0)[r]{\strut{}MPI}}%
      \csname LTb\endcsname
      \put(2084,3732){\makebox(0,0)[r]{\strut{}HPX}}%
      \csname LTb\endcsname
      \put(3527,3732){\makebox(0,0)[r]{\strut{}ITO}}%
      \csname LTb\endcsname
      \put(4970,3732){\makebox(0,0)[r]{\strut{}FBC}}%
      \csname LTb\endcsname
      \put(-356,1983){\rotatebox{-270.00}{\makebox(0,0){\strut{}Efficiency}}}%
      \put(2975,37){\makebox(0,0){\strut{}Number of Nodes}}%
    }%
    \gplbacktext
    \put(0,0){\includegraphics[width={297.60bp},height={198.40bp}]{ata1}}%
    \gplfronttext
  \end{picture}%
\endgroup

%% file: figs/ata16.tex
\begingroup
\Large
  \makeatletter
  \providecommand\color[2][]{%
    \GenericError{(gnuplot) \space\space\space\@spaces}{%
      Package color not loaded in conjunction with
      terminal option `colourtext'%
    }{See the gnuplot documentation for explanation.%
    }{Either use 'blacktext' in gnuplot or load the package
      color.sty in LaTeX.}%
    \renewcommand\color[2][]{}%
  }%
  \providecommand\includegraphics[2][]{%
    \GenericError{(gnuplot) \space\space\space\@spaces}{%
      Package graphicx or graphics not loaded%
    }{See the gnuplot documentation for explanation.%
    }{The gnuplot epslatex terminal needs graphicx.sty or graphics.sty.}%
    \renewcommand\includegraphics[2][]{}%
  }%
  \providecommand\rotatebox[2]{#2}%
  \@ifundefined{ifGPcolor}{%
    \newif\ifGPcolor
    \GPcolortrue
  }{}%
  \@ifundefined{ifGPblacktext}{%
    \newif\ifGPblacktext
    \GPblacktextfalse
  }{}%
  \let\gplgaddtomacro\g@addto@macro
  \gdef\gplbacktext{}%
  \gdef\gplfronttext{}%
  \makeatother
  \ifGPblacktext
    \def\colorrgb#1{}%
    \def\colorgray#1{}%
  \else
    \ifGPcolor
      \def\colorrgb#1{\color[rgb]{#1}}%
      \def\colorgray#1{\color[gray]{#1}}%
      \expandafter\def\csname LTw\endcsname{\color{white}}%
      \expandafter\def\csname LTb\endcsname{\color{black}}%
      \expandafter\def\csname LTa\endcsname{\color{black}}%
      \expandafter\def\csname LT0\endcsname{\color[rgb]{1,0,0}}%
      \expandafter\def\csname LT1\endcsname{\color[rgb]{0,1,0}}%
      \expandafter\def\csname LT2\endcsname{\color[rgb]{0,0,1}}%
      \expandafter\def\csname LT3\endcsname{\color[rgb]{1,0,1}}%
      \expandafter\def\csname LT4\endcsname{\color[rgb]{0,1,1}}%
      \expandafter\def\csname LT5\endcsname{\color[rgb]{1,1,0}}%
      \expandafter\def\csname LT6\endcsname{\color[rgb]{0,0,0}}%
      \expandafter\def\csname LT7\endcsname{\color[rgb]{1,0.3,0}}%
      \expandafter\def\csname LT8\endcsname{\color[rgb]{0.5,0.5,0.5}}%
    \else
      \def\colorrgb#1{\color{black}}%
      \def\colorgray#1{\color[gray]{#1}}%
      \expandafter\def\csname LTw\endcsname{\color{white}}%
      \expandafter\def\csname LTb\endcsname{\color{black}}%
      \expandafter\def\csname LTa\endcsname{\color{black}}%
      \expandafter\def\csname LT0\endcsname{\color{black}}%
      \expandafter\def\csname LT1\endcsname{\color{black}}%
      \expandafter\def\csname LT2\endcsname{\color{black}}%
      \expandafter\def\csname LT3\endcsname{\color{black}}%
      \expandafter\def\csname LT4\endcsname{\color{black}}%
      \expandafter\def\csname LT5\endcsname{\color{black}}%
      \expandafter\def\csname LT6\endcsname{\color{black}}%
      \expandafter\def\csname LT7\endcsname{\color{black}}%
      \expandafter\def\csname LT8\endcsname{\color{black}}%
    \fi
  \fi
    \setlength{\unitlength}{0.0500bp}%
    \ifx\gptboxheight\undefined%
      \newlength{\gptboxheight}%
      \newlength{\gptboxwidth}%
      \newsavebox{\gptboxtext}%
    \fi%
    \setlength{\fboxrule}{0.5pt}%
    \setlength{\fboxsep}{1pt}%
    \definecolor{tbcol}{rgb}{1,1,1}%
\begin{picture}(5952.00,3968.00)%
    \gplgaddtomacro\gplbacktext{%
      \csname LTb\endcsname
      \put(552,612){\makebox(0,0)[r]{\strut{}0\%}}%
      \csname LTb\endcsname
      \put(552,1298){\makebox(0,0)[r]{\strut{}25\%}}%
      \csname LTb\endcsname
      \put(552,1984){\makebox(0,0)[r]{\strut{}50\%}}%
      \csname LTb\endcsname
      \put(552,2669){\makebox(0,0)[r]{\strut{}75\%}}%
      \csname LTb\endcsname
      \put(552,3355){\makebox(0,0)[r]{\strut{}100\%}}%
      \csname LTb\endcsname
      \put(690,382){\makebox(0,0){\strut{}$1$}}%
      \csname LTb\endcsname
      \put(2214,382){\makebox(0,0){\strut{}$2$}}%
      \csname LTb\endcsname
      \put(3737,382){\makebox(0,0){\strut{}$4$}}%
      \csname LTb\endcsname
      \put(5261,382){\makebox(0,0){\strut{}$8$}}%
    }%
    \gplgaddtomacro\gplfronttext{%
      \csname LTb\endcsname
      \put(641,3732){\makebox(0,0)[r]{\strut{}MPI}}%
      \csname LTb\endcsname
      \put(2084,3732){\makebox(0,0)[r]{\strut{}HPX}}%
      \csname LTb\endcsname
      \put(3527,3732){\makebox(0,0)[r]{\strut{}ITO}}%
      \csname LTb\endcsname
      \put(4970,3732){\makebox(0,0)[r]{\strut{}FBC}}%
      \csname LTb\endcsname
      \put(-356,1983){\rotatebox{-270.00}{\makebox(0,0){\strut{}Efficiency}}}%
      \put(2975,37){\makebox(0,0){\strut{}Number of Nodes}}%
    }%
    \gplbacktext
    \put(0,0){\includegraphics[width={297.60bp},height={198.40bp}]{ata16}}%
    \gplfronttext
  \end{picture}%
\endgroup

%% file: 05relatedwork.tex
\section{Related Work}\label{sec:06-Related-Work}

Parallel runtimes are traditionally evaluated using proxy applications such as LULESH~\cite{lulesh} or TeaLeaf~\cite{teaLeaf}, which require significant implementation effort.
Task Bench~\cite{TaskBench} addresses this limitation by providing a parameterized framework for generic task graph execution, previously used to evaluate systems such as Legion, Realm, StarPU, and HPX~\cite{TaskBenchCharmHPX,HoqueRuntimeMicrobench18}.
While earlier evaluations of Itoyori~\cite{Itoyori23} focused on specific algorithms (e.g., Cilksort and ExaFMM~\cite{exaFMM}) and comparisons against MassiveThreads~\cite{massiveThreads}, direct comparisons with other established systems have been lacking.
This work addresses that gap by integrating Itoyori and ItoyoriFBC into Task Bench and comparing them against HPX and MPI.

In the broader landscape of cluster-based parallel programming, UPC++~\cite{UPCxx19} provides a high-performance PGAS communication framework with asynchronous remote procedure calls, offering an alternative RDMA-based approach to distributed memory parallelism.
OmpSs-2@Cluster~\cite{OmpSs2Cluster22} extends the OmpSs task model to distributed memory, enabling task parallelism across cluster nodes.
These systems represent alternative architectural strategies for distributed memory parallelism, distinct from the NFJ model of Itoyori.

Regarding programmer productivity, while the benefits of PGAS~\cite{pgasProd} and AMT~\cite{TaxoTasksPPAM17} are widely recognized, quantitative evaluation remains challenging.
Studies often rely on user surveys~\cite{asc} or software metrics~\cite{productivity}.
This work adopts the latter approach, using LOC and NLC metrics to objectively assess productivity.

In addition to load balancing, modern AMT runtimes are expected to support features such as fault tolerance and dynamic resource management.
Task-level fault tolerance~\cite{PosnerFT20} has been shown to efficiently recover from node failures via checkpointing while avoiding restarting with shrinking recovery.
Task-level dynamic resource management allows AMT applications to grow or shrink dynamically at runtime, thereby improving overall supercomputer utilization~\cite{PosnerDPP25}.
Although this study focuses on load balancing overhead evaluation, extending Task Bench to assess fault tolerance and dynamic resource management capabilities represents a promising direction for future AMT runtime comparisons.

%% file: 06conclusions.tex
\section{Conclusions}\label{sec:07-Conclusions}

We presented the first Task Bench implementations of the Itoyori and ItoyoriFBC runtime systems and compared them with MPI and HPX across diverse scenarios, identifying distinct performance and productivity trade-offs.

The Itoyori implementation demonstrates that the PGAS model combined with global fork-join parallelism can achieve high efficiency in communication-intensive workloads, outperforming the MPI implementation in all-to-all patterns.
Its global address space also significantly improves programmer productivity, reducing code size by nearly 50\% compared to the MPI implementation.
However, its random work stealing scheduler incurs overheads at low task granularities, limiting effectiveness for fine-grained, regular workloads.

The ItoyoriFBC implementation introduces a flexible future-based model supporting arbitrary task graphs.
While its current implementation incurs overheads from explicit task spawning and future management, it offers a promising direction for irregular applications not easily expressed with NFJ structures.

Moreover, we improved the HPX Task Bench implementation.
Results indicate that the HPX implementation excels at intra-node load balancing but faces scalability challenges in communication-heavy scenarios.

Future research should investigate an improved ItoyoriFBC implementation that does not require barriers or future re-wrapping.
Additionally, alternative HPX implementations using nested fork-join or future-based paradigms could help isolate the impact of programming style from runtime characteristics.
Strong scaling analysis also remains relevant future work.
In this work, we used AMT’s default work-stealing algorithm and parameter settings.
A sensitivity study of these parameters is needed to analyze their influence on performance.